\documentclass[Times1COL]{WileyNJDv5}

\articletype{Article Type}%

\received{Date Month Year}
\revised{Date Month Year}
\accepted{Date Month Year}
\journal{Journal}
\volume{00}
\copyyear{2023}
\startpage{1}

\usepackage{amsthm}
\usepackage{array}
\usepackage{color}
\usepackage{rotating}
\usepackage{eurosym}
\usepackage{cancel}
\usepackage{subfigure}

\begin{document}

\title{Risk-Adjusted learning curve assessment using comparative probability metrics}

\author[1]{Adel Ahmadi Nadi}

\author[1]{Stefan H. Steiner}

\author[1]{Nathaniel T. Stevens}

\authormark{Ahmadi Nadi \textsc{et al.}}
\titlemark{}

\address[]{\orgdiv{Department of Statistics and Actuarial Science}, \orgname{University of Waterloo}, \orgaddress{\state{Waterloo}, \country{Canada}}}

\corres{Corresponding author Adel Ahmadi Nadi \email{adel.ahmadinadi@uwaterloo.ca}}



\abstract[Abstract]{Surgical learning curves are graphical tools used to evaluate a trainee's progress in the early stages of their career and determine whether they have achieved proficiency after completing a specified number of surgeries. Cumulative sum (CUSUM)--based techniques are commonly used to assess learning curves due to their simplicity, but they face criticism for relying on fixed performance thresholds and lacking interpretability. This paper introduces a risk-adjusted surgical learning curve assessment (SLCA) method that focuses on estimation rather than hypothesis testing (which is characteristic of CUSUM-type methods). The proposed method is specifically designed to accommodate right-skewed outcomes, such as surgery durations, which are well-characterized by the Weibull distribution. To evaluate the learning process, the proposed SLCA approach sequentially estimates comparative probability metrics that assess the likelihood of a clinically important difference between the trainee's performance and a standard performance. Given the expectation that a trainee's performance will improve over time, we employ a weighted estimating equations approach to the estimation framework to assign greater weight to more recent outcomes compared to earlier ones. Compared to CUSUM-based methods, the proposed methodology offers enhanced interpretability and deeper insights. It also avoids reliance on externally defined performance levels, which are often difficult to determine in practice, and it emphasizes assessing clinical equivalence or noninferiority rather than simply identifying a lack of difference. The effectiveness of the proposed method is demonstrated through a case study on a colorectal surgery dataset as well as a numerical study.}

\keywords{Learning curve, operative time, probability of agreement, weighted estimation equation, Weibull regression}


\maketitle

\renewcommand\thefootnote{}
\footnotetext{\textbf{Abbreviations:} }

\renewcommand\thefootnote{\fnsymbol{footnote}}
\setcounter{footnote}{1}
 \section{Introduction}

The concept of a learning curve was first introduced in aircraft manufacturing to evaluate the impact of increasing production on the time required to produce a single item \cite{wright1936factors}. Since then, learning curve methods have found diverse applications in clinical studies (see, e.g., \cite{harrysson2014systematic, valsamis2018learning, ahn2021learning, burghgraef2022learning, patel2023scoping}). In the setting of surgical learning curves (SLC), the concept is used to evaluate a trainee’s learning process during the early stages of performing a specific surgery. SLCs are typically represented graphically, with time or the number of cases (as a proxy for experience) on the x-axis and a performance measure on the y-axis. The performance measure is typically a function of a surgical outcome (e.g., operative time or 30-day mortality). The outcome of interest is often a clinically relevant characteristic that can vary in nature (e.g., binary, continuous, categorical) depending on the specific surgery being assessed. An SLC can be designed to address various questions; however, the most critical question it seeks to answer is whether the surgeon has gained sufficient experience or training to achieve proficiency within a specific time frame.

CUSUM-type control charts are commonly used to monitor health-related parameters \cite{woodall2015monitoring, lim2002assessing, knoth2019risk, gomon2024inspecting}. These charts are also frequently applied to SLCs to assess the impact of the learning process on the quality of a performed procedure \cite{woodall2021overview}. The simplest and most common CUSUM-based SLC approach plots the monitoring statistic  $C_i = \sum_{j=1}^i (Y_j - \mu_0)$, where $Y_j$ is the outcome for the $j$-th patient  (often with smaller values being preferred), and $\mu_0$ denotes an adequate performance level \cite{forbes2007learning, forbes2004cumulative, forbes2007association}. Let $\mu$ represent the (actual) mean operative time (though it could denote the average of \textit{any} surgical outcome), and then define $\mu_0$ and $\mu_1$ to be predefined adequate and inadequate performance levels, respectively. The CUSUM method seeks to test a null hypothesis $(H_0)$ that the surgeon's performance is adequate versus an alternative hypothesis $(H_1)$ that the performance is inadequate. Biau et al. \cite{biau2008quantitative, biau2010method} introduced the LC-CUSUM, which reverses the hypotheses of the traditional CUSUM approach. This reversal accounts for the reality that trainees typically start with inadequate performance and improve with experience. Thus, the LC-CUSUM signal indicates that the trainee has achieved adequate performance. Since its introduction, the LC-CUSUM has been widely applied across various surgeries, as seen in studies such as \cite{ward2014analysis, dong2021comprehensive, eisenberg2017applying}. For recent advancements in learning curve approaches for binary data, readers are referred to \cite{govindarajulu2018survival, wu2022analyzing}.

Generally, the literature on SLCs for continuous outcomes is thin compared to the binary scenario. However, there are many SLC applications where continuous surgical outcomes are more interesting and more informative than binary ones. Examples of such continuous outcomes include operative time measured in minutes or hours \cite{turan2015season, jimenez2013learning}, length of hospital stay in days \cite{gezer2016cusum}, and intraoperative blood loss measured in millilitres \cite{van2016outcome}. The most commonly used CUSUM-type SLC technique for continuous outcomes, particularly for operative time (OT), is defined by the statistic $C_i = \sum_{j=1}^i (Y_j - \overline{Y}),\, i=1, 2, \dots, t,$ where $\overline{Y} = \frac{1}{t} \sum_{i=1}^t Y_i$  is the sample mean of all $t$ observed values \cite{ward2014analysis, wang2016learning, kim2021evaluation, dimitrovska2022learning, lin2023cusum}. Woodall et al. \cite{woodall2021overview} and Rakovich et al. \cite{rakovich2022comment} strongly advise against using this method for continuous outcomes, as it can be misleading and is often misinterpreted. This is primarily because it substitutes the external standard $\mu_0$ with the sample average $\overline{Y}$, which can lead to incorrect assessments. Additionally, it is important to note that this approach can only be applied retrospectively, requiring the entire dataset of size $t$ to be collected beforehand.

Surgical datasets often include patient-level risk factors (e.g., age, body mass index) that can influence surgery outcomes, necessitating methodological adjustments to account for these factors \cite{steiner2014risk}. In the statistical process monitoring (SPM) area, Steiner et al. \cite{steiner2000monitoring, steiner2001risk} developed risk-adjusted control charts to monitor patient mortality rates. In the context of SLCs, risk adjustment enhances performance assessment by isolating the trainee's contribution through separate estimation of patient characteristics' effects. The traditional CUSUM statistic $C_i = \sum_{j=1}^i (Y_j - \mu_0)$ may fail to provide fair conclusions in such situations as the actual outcome $Y_j$ is influenced by risk factors, while $\mu_0$ ignores them. Therefore, risk adjustment to the benchmark $\mu_0$ is necessary to fairly assess the learning process and ensure that conclusions drawn are reflective of the surgeon's performance, independent of patient characteristics.

Choosing a clinically relevant surgical outcome to quantify performance has been a critical point of discussion among researchers in the field \cite{khan2014measuring}. Several types of variables have been explored in the SLC literature to measure trainee performance. Among these, binary outcomes (e.g., 30-day mortality) and continuous outcomes (e.g., OT) are commonly used. We develop methods for use when OT is the outcome of interest, as it has been demonstrated to be an important quality indicator of a surgeon's performance, given its associations with a wide range of postoperative complications \cite{inaba2019operative}. Research indicates that shorter OTs are associated with improved patient outcomes, whereas longer OT can adversely affect critical outcomes, including higher complication rates and increased financial costs \cite{uecker2013comparable, jackson2011does, kim2014operative, lane2020correlation, offodile2017impact}. Given these implications, OT is a relevant and valuable measure for assessing surgical performance.

Since the OT measures surgery duration, its distribution is typically right-skewed rather than symmetric. Consequently, statistical models such as the Weibull and log-normal distributions are more suitable for modelling OT than the normal distribution, as demonstrated in datasets analyzed by \cite{maggino2018impact, joustra2013can}. Among available statistical models, the Weibull distribution is particularly valuable due to its flexibility; through different specifications of its shape and rate parameters, it can model a broad spectrum of real-world data. Additionally, the Weibull model has been shown to approximate other parametric distributions commonly used in survival and reliability analysis, such as the gamma and log-normal distributions \cite{balasooriya1994selecting, carroll2003use}.  This flexibility makes the Weibull distribution a reasonable choice for modelling OT in clinical studies, as evidenced by its use in studies by \cite{ying2008weibull, zhang2012modeling, glance2018variability}. Weibull regression is also recognized as a good candidate for effectively adjusting for risk factors in reliability and survival analyses \cite{stevens2020bayesian}.

While the CUSUM-type SLC approach offers advantages such as ease of design and implementation, it has been subject to several criticisms. For instance, the LC-CUSUM approach requires specifying $\mu_0$ and $\mu_1$. Although universally accepted and validated performance standards are desirable, many authors have highlighted the lack of such standards for numerous surgical procedures \cite{lim2002assessing, biau2010method}. As a result, most existing studies suggest that these standards be selected through consensus among local surgeons or specialists. On the other hand, it is well-known from the SPM literature that the performance of the CUSUM chart highly depends on the difference between $\mu_0$ and $\mu_1$, so in the absence of meaningful specification of these values, the chart performance is questionable. Consequently, the selection of these standards can also influence the performance of SLC techniques.

 While there have been significant advancements in risk-adjusted SLC approaches, many practical issues remain, especially when dealing with continuous outcomes. Challenges such as building the risk model and estimating its parameters have not been adequately addressed. This means that the existing CUSUM-based SLC approaches for binary variables are often hard to apply to continuous data. For example, while Biau and Porcher \cite{biau2010method} provided details on the risk-adjusted version of the LC-CUSUM approach, the case of continuous outcomes was left undiscussed.

In the present paper, we introduce a risk-adjusted surgical learning curve assessment (SLCA) method that focuses on estimation rather than hypothesis testing (what all CUSUM-type approaches are rooted in). Suppose we are interested in assessing the performance of a trainee relative to a standard performance when performing a specific procedure, where OT represents the outcome of interest. This standard performance could reflect the performance of an established surgeon, a group of surgeons, or an institution. Using comparative probability metrics (CPM), such as the probability of agreement or noninferiority \cite{stevens2017assessing}, we develop an SLCA method that allows us to adjust for risk factors while accounting for the context of the comparison. These metrics assess the likelihood of a clinically important difference between the trainee's performance and the standard one. The proposed method quantifies the impact of experience by updating the estimate of the CPM successively over time. This estimation accounts for the effect of experience through a weighted estimating equations (WEE) approach proposed by Steiner and Mackay \cite{steiner2014monitoring}. As a new surgeon’s skill improves over time, the operative time, controlling for risk factors, is likely to decrease from a relatively high level initially to a lower level as experience is gained. Thus, the WEE approach gives more weight to the recent data, and CPM estimates are updated prospectively as new surgical data is collected.

Based on the above discussion, some advantages of the CPM-based methodology we propose compared to the LC-CUSUM approach—the most commonly used alternative in the SLC literature—are as follows.  The LC-CUSUM method focuses primarily on the end result of the learning process—whether competency has been achieved or not—without addressing how that competency was developed. In contrast, our proposed method offers advantages in terms of interpretability and the depth of information provided over time and across varying patient characteristics. Unlike the LC-CUSUM, the proposed method provides a comprehensive evaluation by estimating multiple performance measures (e.g., CPM and mean operative time) throughout the learning process. This continuous tracking of interpretable performance measures, enhanced by various visualization tools, allows clinicians to gain a deeper understanding of how the trainee’s skills evolve over time and across patients with different characteristics. 

On the other hand, the proposed method does not rely on externally defined performance levels $\mu_0$ and $\mu_1$, which are often difficult to establish in practical situations. Instead, the proposed SLCA approach is based on a known statistical model that describes standard performance, which can be easily estimated using available historical data—a resource commonly accessible in the healthcare sector.
 
 Another advantage of the proposed method over the LC-CUSUM method lies in its emphasis on assessing clinical equivalence or noninferiority rather than merely establishing non-difference. While the LC-CUSUM approach seeks sufficient evidence to confirm that the trainee's performance matches a fixed adequate level, the proposed method measures the likelihood that the trainee’s performance falls within a clinically acceptable range around the standard level. Focusing on equivalence and noninferiority is relevant and more realistic, where replicating exact standards can be challenging for trainees early on, while performing within an acceptable range of those standards may be sufficient.

To the best of our knowledge, a risk-adjusted version of the LC-CUSUM approach has not been explored in the SLC literature for continuous surgical outcomes. This absence of risk adjustment limits the method's applicability in diverse patient populations, as it fails to account for how competency might vary across patients with different characteristics. In contrast, the proposed method incorporates both risk factors and experience into the analysis, enabling us to answer a more specific and clinically relevant question: Has the surgeon gained enough experience to operate safely on patients with specific conditions?

The remainder of this paper is organized as follows. Section 2 presents the proposed methodology, covering the risk-adjusted operative time model, the CPMs and their interpretation, the weighted estimation framework, and the proposed SLCA approach. A case study involving colorectal surgery data is presented in Section 3. Section 4 presents a comparison study that evaluates the performance of the proposed method against the LC-CUSUM method.  Finally, Section 5 provides a discussion and conclusion based on the findings of this study.

\section{The proposed methodology}

This section outlines the proposed methodology, beginning with the risk-adjusted operative time model and the CPM employed. We then introduce the WEE approach and conclude with the implementation procedure of the proposed SLCA method along with some practical recommendations.
 
\subsection{Risk-adjusted outcome model}

Let us assume that a trainee has begun performing a specific type of surgery, and data on $t$ patients have been collected so far. The corresponding operating times are denoted by $Y_i$, with smaller values considered better. For patient $i = 1, 2, \ldots, t$, a total of $d$ risk factors (covariates) are measured prior to the surgery, stored in the vector $\mathbf{x}_i = (x_{i1}, x_{i2}, \ldots, x_{id})$, and these are expected to influence the outcome $Y_i$.  Throughout this paper, the index $i$ will interchangeably represent time, case number, or experience level. Additionally, we assume that a standard performance is defined and available in advance. This standard performance could reflect the performance of an established surgeon, a group of surgeons, or an institution, and it could be estimated with negligible error using historical data that is commonly available in the healthcare sector.

Let us assume that $Y$ follows a Weibull distribution with rate parameter  $\theta>0$ and shape parameter $\eta >0$ and the probability density function
\begin{align}\label{equ:pdfweibull}
f(y)&= \theta \eta y^{\eta-1} e^{-\theta y^{\eta}}.
\end{align}
To account for risk factors, the Weibull regression approach suggests keeping the value $\eta$ constant and adjusting the rate parameter $\theta$ for  $\mathbf{x}$ through a regression model \cite{stevens2020bayesian}. We do this by defining the rate parameter as $\theta=\gamma e^{\pmb{\beta}^\top\textbf{x}}$ so that $\pmb{\beta}^\top\textbf{x}=\beta_1 x_1 + \beta_2 x_2 + \cdots + \beta_d x_d$. By this assumption, the conditional probability density function of $Y$ given $\mathbf{x}$ under the Weibull regression model is
\begin{align}\label{equ:pdfweibull}
f(y|\textbf{x})=\gamma e^{\pmb{\beta}^\top\textbf{x}} \eta y^{\eta-1} e^{- \gamma e^{\pmb{\beta}^\top\textbf{x}} y^{\eta}}.
\end{align}
Using \eqref{equ:pdfweibull}, the risk-adjusted mean operative time (RMOT) can be calculated as
\begin{align}\label{equ:meanweibull}
\mu(\textbf{x})=E(Y|\textbf{x})=\frac{\Gamma(\frac{1}{\eta}+1)}{\sqrt[\eta]{\theta}}=\frac{\Gamma(\frac{1}{\eta}+1)}{\sqrt[\eta]{\gamma}} e^{\frac{-1}{\eta}\pmb{\beta}^\top\textbf{x}},
\end{align}
where $\Gamma(\cdot)$ is the complete gamma function. According to \eqref{equ:meanweibull}, risk factors have a multiplicative effect on the RMOT so that $\beta_{k}$ for $k=1, 2, \ldots,d$ quantifies the change in magnitude of RMOT per unit change in the risk factor $x_{k}$. In particular,  a positive (negative) $\beta$ value suggests that an increase in the risk factor is associated with a decrease (increase) in operative time.

\subsection{Estimating the parameters using the WEE approach}
 \label{sec:estim}

 To apply the proposed SLCA approach, which will be detailed in Subsection \ref{sec:SLC}, it is necessary to estimate the unknown parameters in \eqref{equ:pdfweibull} and calculate the sampling distribution of their estimators. Before proceeding, we emphasize that since the standard performance (modelled by a risk-adjusted Weibull model with parameters $\gamma_S$, $\eta_S$, and $\pmb{\beta}_S$) is assumed to be known, this section will focus only on estimating the trainee's model parameters $\gamma_N$, $\eta_N$, and $\pmb{\beta}_N$. Since estimation is only relevant for the trainee, in this section, we will occasionally omit the subscript $N$ for brevity. The subscript $S$ will be retained, however, to emphasize when we are discussing the standard performance. In Section \ref{sec:SLC}, we discuss an adaptation of our methodology that incorporates estimation of $\gamma_S, \eta_S,$ and $\beta_S$.

  In what follows, we explain the maximum likelihood estimation method used to estimate the parameters, which incorporates the WEE approach proposed by Steiner and Mackay \cite{steiner2014monitoring}. Let us denote the vectors of all available OT data and the corresponding vector of risk factors by $\textbf{y}=(y_1, y_2, \ldots, y_t)$ and $\textbf{x}=(\textbf{x}_1, \textbf{x}_2, \ldots, \textbf{x}_t)$, respectively.  To estimate the parameters based on the maximum likelihood method,  the log-likelihood function
 \begin{align}\label{equ:LLF}
\ell\left(\gamma, \eta, \pmb{\beta} | \textbf{y}, \textbf{x}\right)=\sum_{i=1}^{t} \ell_i \left(\gamma, \eta, \pmb{\beta} | y_{i}, \textbf{x}_{i}\right)
 \end{align}
needs to be maximized with respect to $\gamma, \eta$ and $\pmb{\beta}$. The $\ell_i$ here is the log-likelihood contribution from the observations $(y_{i}, \textbf{x}_i)$ based on the $i$-th surgery and calculated from the conditional probability density function in \eqref{equ:pdfweibull} as
  \begin{align}\label{equ:llf}
\ell_{i}\left(\gamma, \eta, \pmb{\beta} | y_{i}, \textbf{x}_{i}\right)=\ln(\gamma)+\ln(\eta)+(\eta -1) \ln(y_{i})+\pmb{\beta}^\top \textbf{x}_{i}-\gamma y_{i}^{\eta} e^{\pmb{\beta}^\top\textbf{x}_{i}}.
 \end{align}

The log-likelihood function in \eqref{equ:LLF} assumes the patients are independent. Furthermore, it assigns the same weight to all data points, from the past to the present. This approach leads to bias in the estimates of present performance, as the performance is expected to change over time. Conversely, using a recent portion of data (instead of all available data) will result in more variation associated with estimation.  By using all available historical data but assigning more weight to the current outcomes (which are more relevant to the current performance of the trainee) than the distant past ones, the WEE technique estimates the current performance with less bias than using the data unweighted and less uncertainty than an analysis based only on very recent data. To accomplish this, we first define the weights and then present the mathematical foundation of the WEE method. We define the patient-specified weight
\begin{align}\label{weight}
w_{i}=\frac{t \lambda(1-\lambda)^{(t-i)}}{[1-(1-\lambda)^{t}]}, \quad \text{for}  \quad i=1, 2, \ldots,t,
\end{align}
where $0<\lambda \leq 1$ is the smoothing constant.  According to \eqref{weight}, values of $\lambda$ close to zero (one) give more (less) weight to observations in the distant past.

To apply the WEE approach in this scenario, the weighted log-likelihood function can be calculated as
 \begin{align} 
\sum_{i=1}^{t} w_{i} \ell_{i}\left(\gamma, \eta, \pmb{\beta} | y_{i}, \textbf{x}_{i}\right),
 \end{align}
where $\ell_{i}$ is given in \eqref{equ:llf}. Let us denote the vector of weights as $\textbf{w}=(w_1, w_2, \ldots,w_t)$. To obtain parameter estimates, $\hat{\gamma}, \hat{\eta},$ and $\hat{\pmb{\beta}}$, we need to simultaneously solve the following system of $(d+2)$ equations 
\begin{align}\label{WEEequ1}
Q\left(\gamma, \eta, \pmb{\beta} | \textbf{y}, \textbf{x}, \textbf{w}\right)=\left[\begin{array}{c}
Q(\gamma | \textbf{y}, \textbf{x}, \textbf{w}) \\
Q(\eta | \textbf{y}, \textbf{x}, \textbf{w}) \\
Q(\beta_{1} | \textbf{y}, \textbf{x}, \textbf{w}) \\
\vdots\\
Q(\beta_{d} | \textbf{y}, \textbf{x}, \textbf{w}) \\
\end{array}\right]=\left[\begin{array}{c}
 \sum_{i=1}^{t} w_{i} \frac{\partial l_{i}\left(\gamma, \eta, \pmb{\beta} | y_{i}, \textbf{x}_{i}\right)}{\partial \gamma} \\
 \sum_{i=1}^{t} w_{i} \frac{\partial l_{i}\left(\gamma, \eta, \pmb{\beta} | y_{i}, \textbf{x}_{i}\right)}{\partial \eta} \\
 \sum_{i=1}^{t} w_{i} \frac{\partial l_{i}\left(\gamma, \eta, \pmb{\beta} | y_{i}, \textbf{x}_{i}\right)}{\partial \beta_{1}} \\
\vdots\\
 \sum_{i=1}^{t} w_{i} \frac{\partial l_{i}\left(\gamma, \eta, \pmb{\beta} | y_{i}, \textbf{x}_{i}\right)}{\partial \beta_{d}} \\
\end{array}\right]=\left[\begin{array}{c}
0 \\
0 \\
0 \\
\vdots\\
0\\
\end{array}\right].
\end{align}

The uncertainty in the parameter estimators, $\tilde{\gamma}, \tilde{\eta},$ and $\tilde{\pmb{\beta}}$, can be quantified by their variance-covariance matrix. To estimate the variance-covariance matrix of the estimators obtained using the WEE approach, we adapted the method used by Steiner and Mackay \cite{steiner2014monitoring}, which was originally introduced by White \cite{white1982maximum}. Given the WEEs in \eqref{WEEequ1},  the variance-covariance matrix of the vector of estimators $(\tilde{\gamma},\tilde{\eta},\tilde{\pmb{\beta}}) \in \mathbb{R}^{d+2}$  has the following sandwich form 
\begin{align}\label{equ:sigmabj}
\pmb{\Sigma}=\left(\pmb{\Gamma}\right)^{-1} \pmb{\Omega} \left(\pmb{\Gamma}\right)^{-1},
\end{align}
where
\begin{align}\label{EHesmat}
 \pmb{\Gamma}=\mathbb{E} \left(\frac{\partial Q\left(\gamma, \eta, \pmb{\beta} | \textbf{Y}, \textbf{x}, \textbf{w}\right)}{\partial(\gamma, \eta,\pmb{\beta})}  \right)_{\gamma=\hat{\gamma}, \eta=\hat{\eta},\pmb{\beta}=\hat{\pmb{\beta}}},
 \end{align}
is the expected Hessian matrix and
\begin{align}\label{varcovmatWEE}
\pmb{\Omega}=\mathbb{E} \left(\big[Q\left(\gamma, \eta, \pmb{\beta} | \textbf{Y}, \textbf{x}, \textbf{w}\right)\big] \times \big[Q\left(\gamma, \eta, \pmb{\beta} | \textbf{Y}, \textbf{x}, \textbf{w}\right)\big]^\top\right)_{\gamma=\hat{\gamma}, \eta=\hat{\eta},\pmb{\beta}=\hat{\pmb{\beta}}},
\end{align}
 is the information matrix for the parameters evaluated at  $\gamma=\hat{\gamma}, \eta=\hat{\eta}, \beta_{1}=\hat{\beta}_{1},\ldots,\beta_{d}=\hat{\beta}_{d},$ and $Q\left(\gamma, \eta, \pmb{\beta} | \textbf{Y}, \textbf{x}, \textbf{w}\right)$ is given in \eqref{WEEequ1}. The closed-form mathematical expressions for the elements of $Q\left(\gamma, \eta, \pmb{\beta} \mid \mathbf{y}, \mathbf{x}, \mathbf{w}\right)$ in \eqref{WEEequ1}, as well as the matrices $\pmb{\Gamma}$ and $\pmb{\Omega}$ in \eqref{EHesmat} and \eqref{varcovmatWEE}, can be found in the Appendices A and B. The variance-covariance matrix of the WEE-based estimators, as given in \eqref{equ:sigmabj}, combined with the Delta method, will be employed in the next subsections to calculate the CPM and also to obtain asymptotic confidence intervals for key performance measures that are functions of the model parameters. 
  
\subsection{Comparative probability metrics}
\label{sec:CPM}

In order to evaluate the performance of the new surgeon relative to the standard performance based on patients with the same condition, we define the relative risk $R(\textbf{x})=\frac{\mu_{N}(\textbf{x})}{\mu_{S}(\textbf{x})}$ where $\mu(\textbf{x})$ is given in \eqref{equ:meanweibull} and subscripts $N$ and $S$ respectively denote the RMOT for the new surgeon and the standard.  The information provided by $R(\textbf{x})$ is valuable in understanding how the OT of the new surgeon compares to the standard performance. If $R(\textbf{x}) = 1$, the trainee's expected performance is at the same level as the standard in terms of operative time. And because smaller RMOT is preferred, $R(\textbf{x})<1\, (>1)$ means the new surgeon performs better (worse) than the standard performance. Denote the estimator of this relative risk by $\tilde{R}(\textbf{x})$. To assess the level of agreement or noninferiority between the trainee's performance and the standard, we utilize the CPM framework introduced by Stevens et al. \cite{stevens2017assessing} to define the relevant metric as follows
\begin{align}\label{equ:CPM}
\text{CPM}_{i}(\textbf{x}|\delta_{L},\delta_{U})&=P\Big(\delta_L < \tilde{R}_{i}(\textbf{x}) < \delta_U | \textbf{x} \Big), \, \text{for} \, i=1, 2, \ldots,t,
\end{align}
where $\tilde{R}_i(\textbf{x})$ is the estimator of the relative risk obtained by applying the WEE approach discussed in Subsection \ref{sec:estim} using the data accumulated up to time $i$, i.e., $(y_{j}, \textbf{x}_j), j=1, 2, \ldots, i$. Consequently, CPM$_i$ represents the most up-to-date assessment of the trainee's performance calculated from the trainee's data collected up to time $i$.

The CPM in \eqref{equ:CPM} is simply the probability that the relative risk estimator at time $i$ is contained within the region $(\delta_L, \delta_U$). The values $\delta_L$ and $\delta_U$ $(0 \leq \delta_L < \delta_U \leq \infty$) can be chosen based on the context of the comparison. We propose choosing them in relation to what percent difference between $\mu_{N}(\textbf{x})$ and $\mu_{S}(\textbf{x})$ is clinically relevant. If a $100 \epsilon \%$ change (increase or decrease) represents the smallest percentage change in mean operative time over the standard performance that would not be of clinical importance, the interval $((1+\epsilon)^{-1},(1+\epsilon))$ which is symmetric around $1$ on the relative scale represents the differences that are clinically negligible. 

In light of this, there are some choices for $(\delta_L,\delta_U)$ that yield especially meaningful interpretations. For example
\begin{itemize}
\item[$\bullet$] If $\delta_L=(1+\epsilon)^{-1}$ and $\delta_U=(1+\epsilon)$, then the CPM in \eqref{equ:CPM} is the probability of clinical equivalence between two performances. This is referred to as the probability of agreement (PA). Large values of PA suggest that the trainee is likely to perform similarly to the standard (with respect to the defined indifference region $(\delta_L,\delta_U))$ on patients with pre-operative conditions given by $\textbf{x}$.
\item[$\bullet$]  If  $\delta_L=0$ and $\delta_U=(1+\epsilon)$, then $\text{CPM}_i(\textbf{x}|0,(1+\epsilon))=P\Big(\tilde{R}_i(\textbf{x}) < (1+\epsilon)| \textbf{x}\Big)$  is the probability that the new surgeon's performance is not clinically worse than the standard performance. We call this the probability of noninferiority (PN).
\end{itemize}

The appropriate selection of $\delta_L$ and $\delta_U$, and hence the resulting interpretation of CPM, depends on the specific context and nature of the problem. It is important to emphasize that all conclusions based on the aforementioned probability metrics are conditioned on the vector of risk factors $\mathbf{x}$. A comprehensive comparison across all patients can be achieved by evaluating the CPM over a representative range of $\mathbf{x}$ values.

It is also worth mentioning that the WEE technique allows us to integrate the time factor into the proposed approach. Specifically, the WEE method enables continuous updating of the CPM as data accumulates, allowing it to capture changes in the trainee's performance over time.  In order to calculate the CPM$_{i}$ in \eqref{equ:CPM}, we need the sampling distribution of $\tilde{R}_{i}(\textbf{x})$. According to the asymptotic normality of WEE-based estimators \cite{wang2004asymptotic}, we have $ \tilde{R}_i(\textbf{x}) \dot{\sim} \mathcal{N}(R_i(\textbf{x}),\sigma^2_{R_i}(\textbf{x}))$ where that the variance term $\sigma^2_R(\textbf{x})$ can be derived by applying the Delta method. We have
\begin{align}
\label{equ:sigmaR}
\sigma^2_{R_i}(\textbf{x})&=\Bigg(\frac{\partial R(\textbf{x})}{\partial \gamma}, \frac{\partial R(\textbf{x})}{\partial \eta},\frac{\partial R(\textbf{x})}{\partial \pmb{\beta}_{N}}\Bigg) \pmb{\Sigma}
\Bigg(\frac{\partial R(\textbf{x})}{\partial \gamma}, \frac{\partial R(\textbf{x})}{\partial \eta},\frac{\partial R(\textbf{x})}{\partial \pmb{\beta}}\Bigg)^\top,
 \end{align}
where $\pmb{\Sigma}$ is the $(d+2) \times (d+2)$ variance-covariance matrix given in \eqref{equ:sigmabj}. Accordingly, given  the asymptotic normality of $\tilde{R}_i(\textbf{x})$, the CPM$_i(\textbf{x}|\delta_L,\delta_U)$ in \eqref{equ:CPM} can be approximated at times $i=1,2,\ldots$ using the following normal probability
\begin{align}\label{equ:CPMN}
\text{CPM}_i(\textbf{x}|\delta_L,\delta_U)\approx \Phi\left(\frac{\delta_U - R_i(\textbf{x})}{\sigma_{R_i}(\textbf{x})}\right)-\Phi\left(\frac{\delta_L - R_i(\textbf{x})}{\sigma_{R_i}(\textbf{x})}\right),
\end{align}
where $\Phi$ is the standard normal cumulative distribution function.

\subsection{The proposed SLCA approach}
\label{sec:SLC}

This section introduces the proposed SLCA approach, building on the methodology outlined in the preceding subsections. To capture changes in the trainee’s performance—specifically seeking evidence of learning—relative to the standard performance, we propose sequentially estimating the CPM$_i$ in \eqref{equ:CPMN} by applying the WEE approach at times $i = 1, 2, \ldots$. At each time point $i$, the parameters $\gamma, \eta,$ and  $\pmb{\beta}$ should first be estimated using the WEE approach discussed in Section \ref{sec:estim}, based on the accumulated data from patients up to and including case $i$. Using these estimates at time $i$, we can calculate the estimated relative risk $\hat{R}_i(\mathbf{x}) = \frac{\hat{\mu}_{Ni}(\mathbf{x})}{\mu_{Si}(\mathbf{x})}$, the estimated variance-covariance matrix $\hat{\pmb{\Sigma}}$, and, consequently, the estimated standard deviation $\hat{\sigma}^2_{R_i}(\mathbf{x})$ of $\tilde{R}_{i}(\mathbf{x})$ (as given in \eqref{equ:sigmaR}). Finally, all these estimates are substituted into the equation \eqref{equ:CPMN} to estimate $\text{CPM}_i(\mathbf{x}|\delta_L, \delta_U)$ based on the data up to time $i$. In this way, the proposed method utilizes $\widehat{\text{CPM}}_i(\mathbf{x}|\delta_L, \delta_U)$ for $i=1, 2, \ldots, t$ to assess the learning process over time.

If consecutively estimated CPM values follow an increasing trend for a given $\mathbf{x}$, it indicates that the learning process for patients with preoperative conditions characterized by the vector $\mathbf{x}$ has occurred. At any time point $i$, and for any risk factor $\mathbf{x}$, if $\widehat{\text{CPM}}_i(\mathbf{x}|\delta_{L}, \delta_{U})$ exceeds a predetermined cut-off value (e.g., $0.95$), it can be concluded that the trainee has acquired sufficient competence to operate on a patient with the given risk factor profile. However, if $\widehat{\text{CPM}}_i(\mathbf{x}|\delta_{L}, \delta_{U})$ falls below the cut-off value, indicating a lack of competency, the proposed method can be repeated after observing new surgical results until the new surgeon achieves an adequate level of performance. In this way, the WEE approach offers an additional advantage: it allows for the assessment of learning both retrospectively and prospectively. When using the method prospectively, the CPM estimate only needs to be updated when a new outcome becomes available.

The determination of whether the trainee has reached an adequate level of performance depends on the cut-off value set for the CPM. Based on the chosen values of $\delta_L$ and $\delta_U$, CPM values close to one indicate that the performances of the trainee and the standard surgeon are clinically equivalent or that the trainee’s performance is clinically non-inferior to the standard. On the other hand, CPM values close to zero suggest that the performances are not equivalent or that the trainee is clinically inferior to the standard. Naturally, the question arises: "How close to one (or zero) is sufficient to draw one conclusion over the other?"  Many studies using similar probability metrics in other contexts, such as measurement systems comparison  \cite{stevens2017assessing} and survival analysis \cite{stevens2020comparing}, recommend a cut-off value of $0.95$ as a reasonable benchmark for CPM. However, depending on the specific context, more lenient cut-off values (e.g., $0.7$ or $0.8$) or more stringent ones (e.g., $0.99$) can be applied. The cut-off value should be determined through consultation with experts.

 To aid interpretation, we further suggest visualizing $\widehat{\text{CPM}}$ against time and/or risk factors. Such visualizations can clarify various aspects of the learning process. Specifically, plotting the point and interval estimates of $\text{CPM}_i(\mathbf{x}|\delta_L, \delta_U)$ against the time index $i$ and/or the risk factor vector $\mathbf{x}$ allows for a clearer interpretation of how experience and patient case mix impact the performance of the new surgeon. Such a plot also ensures estimation uncertainty is considered when drawing conclusions. Given the asymptotic normality of the WEE estimators, the $100(1-\alpha)\%$ pointwise asymptotic confidence intervals (ACIs) for $\text{CPM}_i(\mathbf{x}|\delta_L, \delta_U)$ can be derived directly. However, this approach may yield confidence limits outside the interval $(0, 1)$, particularly when $\widehat{\text{CPM}}_i$ is close to $0$ or $1$. To address this issue, we employ an alternative method by constructing ACIs for the transformed quantity $\Psi_i = \ln(-\ln(\text{CPM}_i))$, which maps $(0, 1)$ to $(-\infty, \infty)$. The ACI for $\Psi_i$ is given by
\begin{align}\label{equ:CB}
\left(L_{\Psi_i}(\mathbf{x}), U_{\Psi_i}(\mathbf{x})\right) = \widehat{\Psi}_i(\mathbf{x}|\delta_L, \delta_U) \pm z_{1-\frac{\alpha}{2}} \hat{\sigma}_{\Psi_i}(\mathbf{x}),
\end{align}
where $z_{1-\frac{\alpha}{2}}$ is the $1-\frac{\alpha}{2}$ percentile of the standard normal distribution, and $\hat{\sigma}_{\Psi_i}(\mathbf{x})$ is the estimated standard deviation of $\tilde{\Psi}_i(\mathbf{x}|\delta_L, \delta_U)$. The estimated standard deviation $\hat{\sigma}_{\Psi_i}(\mathbf{x})$ can be computed as
\begin{align}\label{equ:sigmacpm}
\hat{\sigma}_{\Psi_i}^2(\mathbf{x}) = \Bigg(\frac{\partial \Psi(\mathbf{x})}{\partial \gamma}, \frac{\partial \Psi(\mathbf{x})}{\partial \eta}, \frac{\partial \Psi(\mathbf{x})}{\partial \pmb{\beta}}\Bigg) 
\pmb{\Sigma} 
\Bigg(\frac{\partial \Psi(\mathbf{x})}{\partial \gamma}, \frac{\partial \Psi(\mathbf{x})}{\partial \eta}, \frac{\partial \Psi(\mathbf{x})}{\partial \pmb{\beta}}\Bigg)^\top,
\end{align}
evaluated at the estimates $\hat{\gamma}$, $\hat{\eta}$, and $\hat{\pmb{\beta}}$. The limits of the ACI in \eqref{equ:CB} are then transformed back using the inverse transformations $L_{C_i}(\mathbf{x}) = \exp\left(-\exp\left(U_{\Psi_i}(\mathbf{x})\right)\right) $ and $ U_{C_i}(\mathbf{x}) = \exp\left(-\exp\left(L_{\Psi_i}(\mathbf{x})\right)\right)$ to obtain the $100(1-\alpha)\%$ ACI for $\text{CPM}_i(\mathbf{x}|\delta_L, \delta_U)$. This method ensures that the resulting confidence limits remain within the interval $(0, 1)$.

Since $\widehat{\text{CPM}}_i(\textbf{x})$ is a function of both time and a potentially high-dimensional covariate $\textbf{x}$, care needs to be taken when considering how to visualize this surface. When there is only one risk factor ($d=1$), the surface is 3-dimensional and can be plotted against time $i$ and the potential range of risk factor $x$. When $d \geq 2$, one approach is to generate multiple 3-dimensional surfaces by holding all of the risk factors but one constant and then plotting contours of $\widehat{\text{CPM}}_i(\textbf{x})$ against time and the only free risk factor. If the analyst only cares about time and aims to track changes over time, the performance plots can be made for a typical patient with common values for the risk factors. 

We further suggest constructing RMOT and relative risk plots by plotting the point and interval estimates of the $\mu_{N}(\textbf{x})$ and the relative risk $R(\textbf{x})$ against time and/or risk factors. These plots can provide helpful insight into the performance evaluation and assessment. To construct these plots, we use the facts that $\tilde{\mu}(\textbf{x}) \dot{\sim} \mathcal{N}\Big(\mu(\textbf{x}),\sigma^2_{\mu}(\textbf{x})\Big)$ and $ \tilde{R}(\textbf{x}) \dot{\sim} \mathcal{N}(R(\textbf{x}),\sigma^2_R(\textbf{x}))$ where the variance $\sigma^2_{\mu}(\textbf{x})$ can be derived by applying the Delta method as
\begin{align}\label{equ:sigmamu}
\sigma^2_{\mu}(\textbf{x})&= \Big(\frac{\partial \mu(\textbf{x})}{\partial \gamma}, \frac{\partial \mu(\textbf{x})}{\partial \eta},\frac{\partial \mu(\textbf{x})}{\partial \pmb{\beta}}\Big) \pmb{\Sigma} \Big(\frac{\partial \mu(\textbf{x})}{\partial \gamma}, \frac{\partial \mu(\textbf{x})}{\partial \eta},\frac{\partial \mu(\textbf{x})}{\partial \pmb{\beta}}\Big)^\top,
 \end{align}
where $\pmb{\Sigma}$ is given in \eqref{equ:sigmabj}.  Accordingly, $100(1-\alpha)\%$ pointwise ACI for $\mu_i(\textbf{x})$ and $R_i(\textbf{x})$ can be derived as
\begin{align} 
\left(L_{\mu_i}(x),U_{\mu_i}(x)\right)= \hat{\mu}_i(\textbf{x}) \pm z_{1-\frac{\alpha}{2}} \hat{\sigma}_{\mu_i}(\textbf{x}), \qquad \left(L_{R_i}(x),U_{R_i}(x)\right)=  \hat{R}_i(\textbf{x}) \pm z_{1-\frac{\alpha}{2}} \hat{\sigma}_{R_i}(\textbf{x}), \quad \text{for} \quad i=1, 2, \ldots.
\end{align}

We illustrate such plots in our example in Section \ref{sec:example}. We conclude this section by noting that, although we assumed that the standard performance parameters are available in advance, the proposed method can be readily extended to relax this assumption and incorporate the uncertainty introduced by their estimation into the learning curve assessment. To achieve this, we first need to calculate the estimated parameters, $\widehat{\gamma}_S$, $\widehat{\eta}_S$, and $\widehat{\pmb{\beta}}_S$, along with their corresponding variance-covariance matrix, $\pmb{\Sigma}_S$, using an unweighted approach such as the maximum likelihood estimation approach. Then, the joint variance-covariance of the vector $(\tilde{\gamma}_N, \tilde{\eta}_N, \tilde{\pmb{\beta}}_N, \tilde{\gamma}_S, \tilde{\eta}_S, \tilde{\pmb{\beta}}_S)$ is of form $
\begin{pmatrix}
\pmb{\Sigma}_N & \textbf{0} \\
\textbf{0} & \pmb{\Sigma}_S
\end{pmatrix}$ where $\textbf{0}$ is a $(d + 2) \times (d + 2)$ zero matrix. Note that the off-diagonal blocks in this matrix are zero matrices, as we assume the trainee's performance and the standard performance are independent, which is a plausible assumption in practice. Finally, the variance terms in \eqref{equ:sigmaR}, \eqref{equ:sigmacpm}, and \eqref{equ:sigmamu} must be adjusted to incorporate the partial derivatives with respect to all the model parameters and the corresponding joint variance-covariance matrix.

\section{Colorectal surgery example}
\label{sec:example}

\begin{figure}[]
 \centering
\subfigure[Histogram of observed OTs.]{%
\resizebox*{7.5cm}{!}{\includegraphics{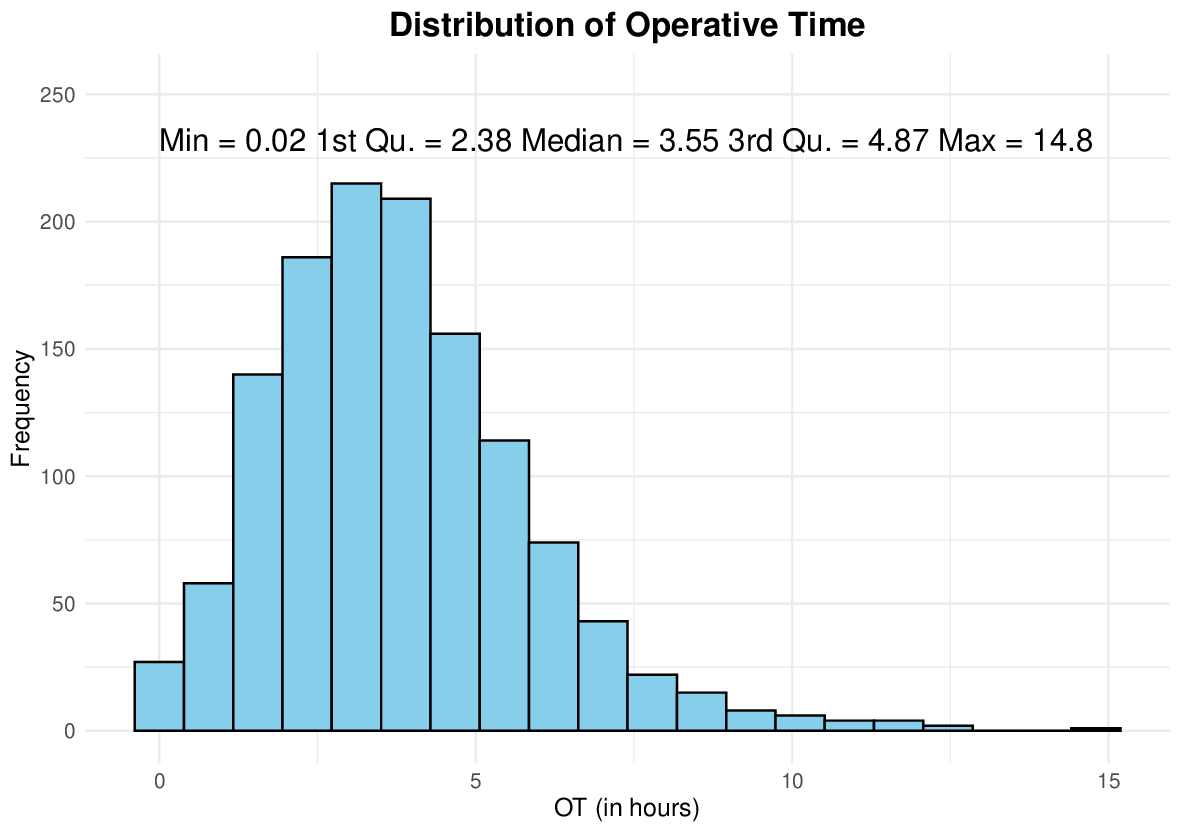}}}
\subfigure[Scatter plot of OTs vs BMIs.]{%
\resizebox*{7.5cm}{!}{\includegraphics{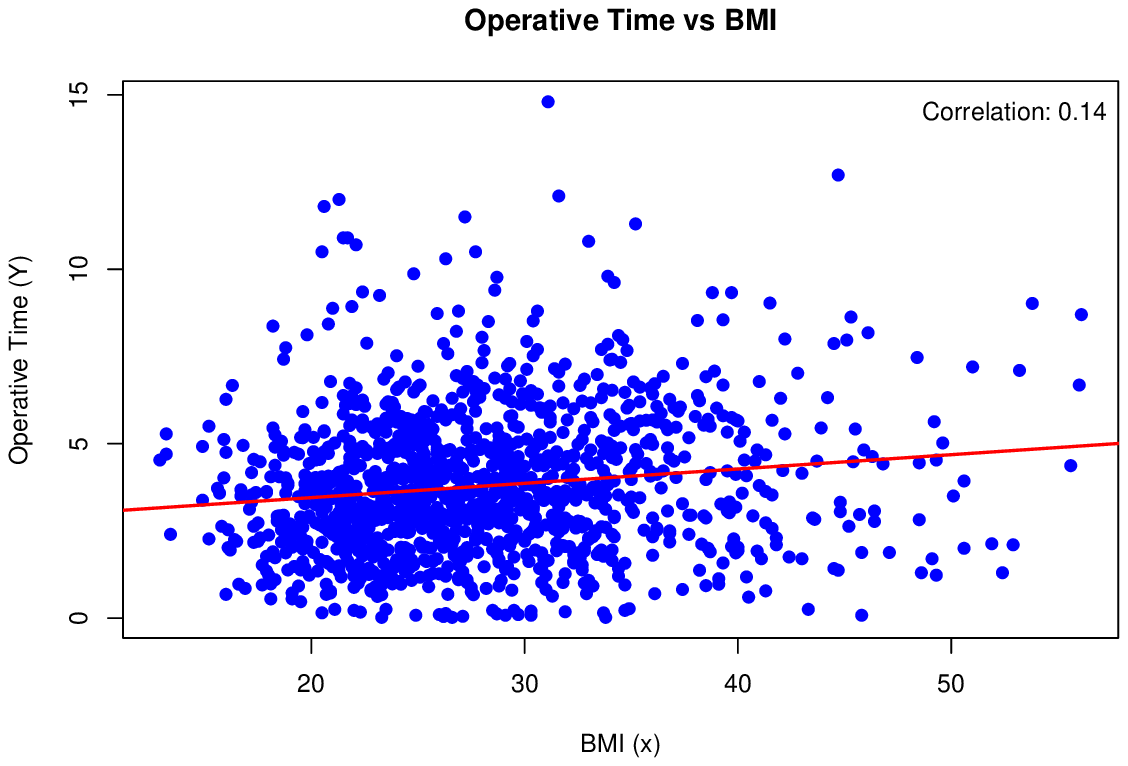}}}
\subfigure[Scatter plot of OTs against the case number (time) with a fitted smoothing line for the initial cohort.]{%
\resizebox*{7.5cm}{!}{\includegraphics{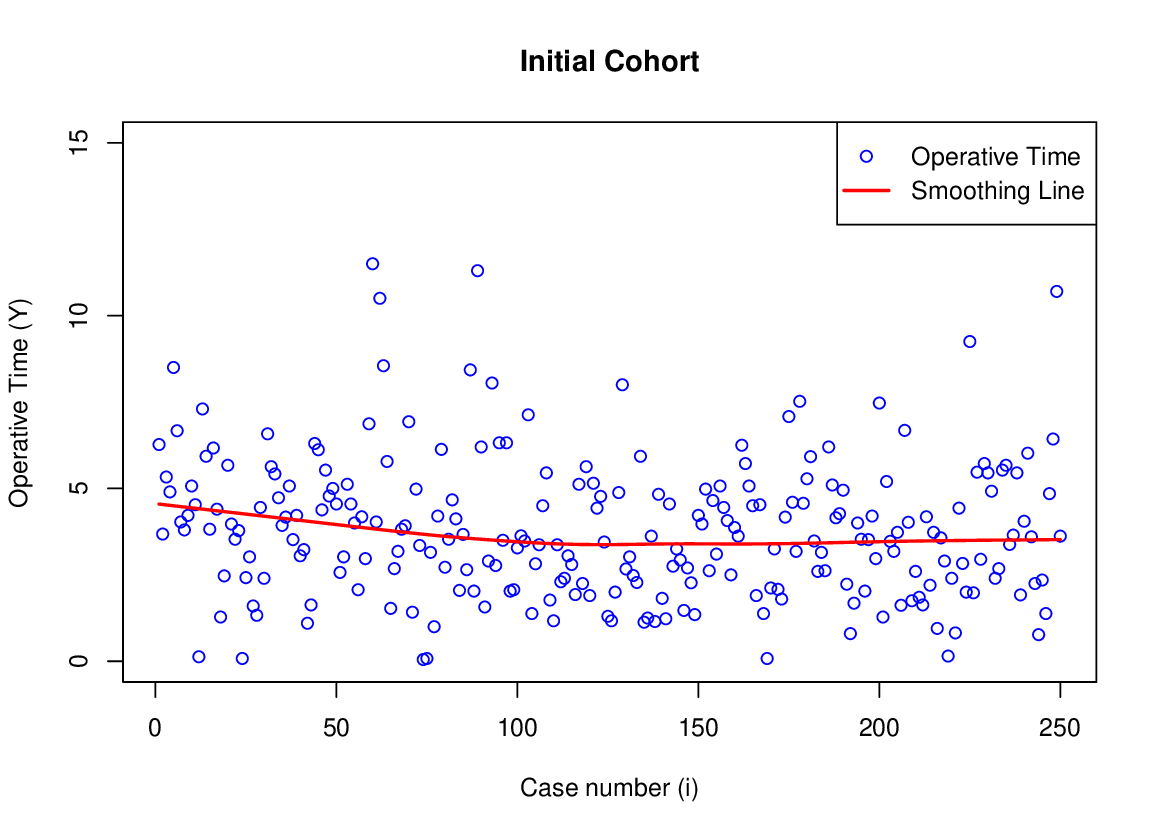}}}
\subfigure[Scatter plot of OTs against the case number (time) with a fitted smoothing line for the standard cohort.]{%
\resizebox*{7.5cm}{!}{\includegraphics{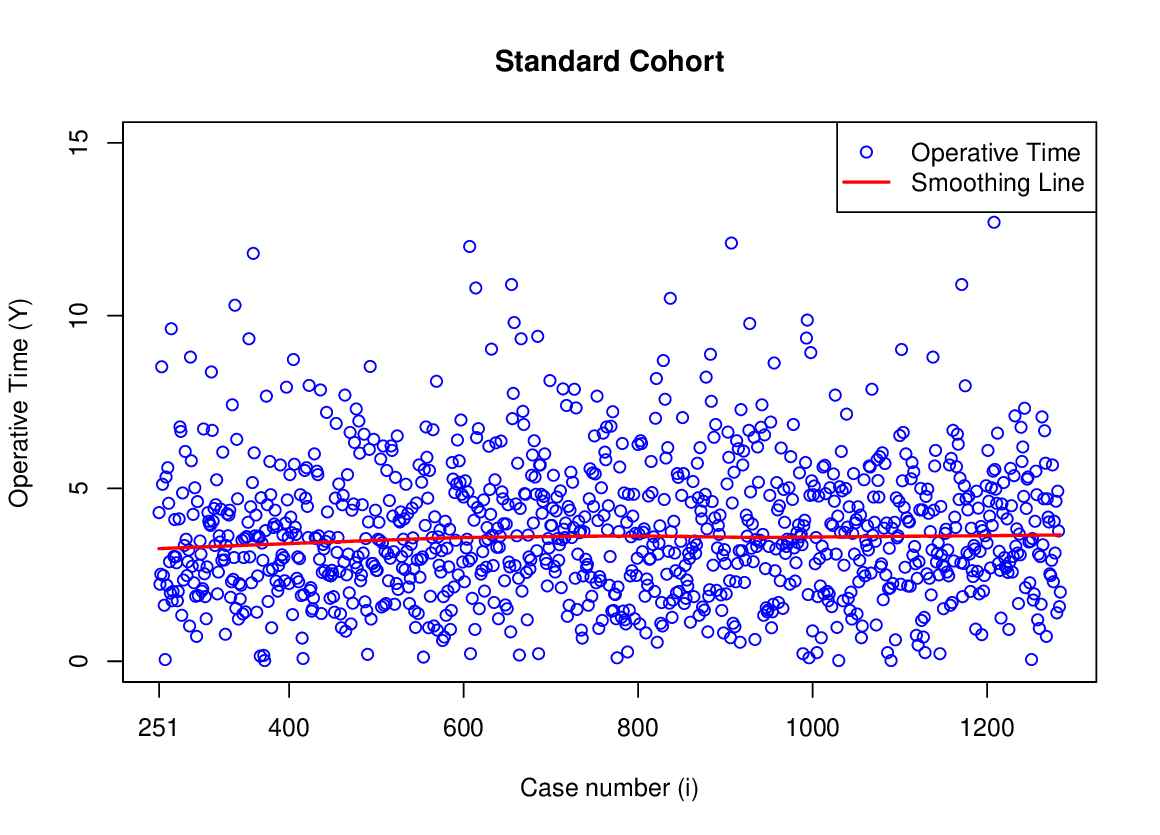}}}
\caption{ Histogram and scatter plots based on raw data.}
\label{fig:RD}
\end{figure}

Colorectal surgery involves procedures to repair damage to the colon, rectum, or anus, often due to cancer, diverticulitis, or inflammatory bowel disease. This type of surgery can be performed using traditional open surgery or minimally invasive techniques such as laparoscopy. In this section, we use the surgical data discussed by Turan et al. \cite{turan2015season} to illustrate our approach. The dataset includes 1,284 patients with severe systemic disease who underwent colorectal surgery at the Cleveland Clinic Main Campus between June 2010 and March 2012. For each patient, in addition to recording OTs (operative times) in hours, body mass index (BMI) was also measured as a risk factor. Figures \ref{fig:RD}(a) and (b) display the histogram of the observed OTs in the dataset and the scatter plot of OTs versus BMIs along with a fitted regression line, respectively. The histogram shows a right-skewed distribution of OTs. The scatter plot indicates a weak positive correlation between OT and BMI, with a correlation value of 0.14. This suggests that OT tends to increase slightly as BMI rises. Unfortunately, the data does not include information about individual surgeons. However, we observed evidence of outcome improvement in the early surgeries, followed by stabilized performance in the later cases. Thus, we use this colorectal surgery dataset to illustrate our approach. To do this, we split the dataset into two cohorts. The initial cohort represents the first 250 surgeries (treated as outcomes under the trainee), and the standard cohort represents the subsequent surgeries (treated as outcomes under the standard performance). Figures \ref{fig:RD}(c) and (d) present scatter plots of OTs against time with a smoothing line for each cohort, supporting the rationale behind this data split.

\begin{figure}[]
 \centering
\subfigure[Standard cohort.]{%
\resizebox*{7.5cm}{!}{\includegraphics{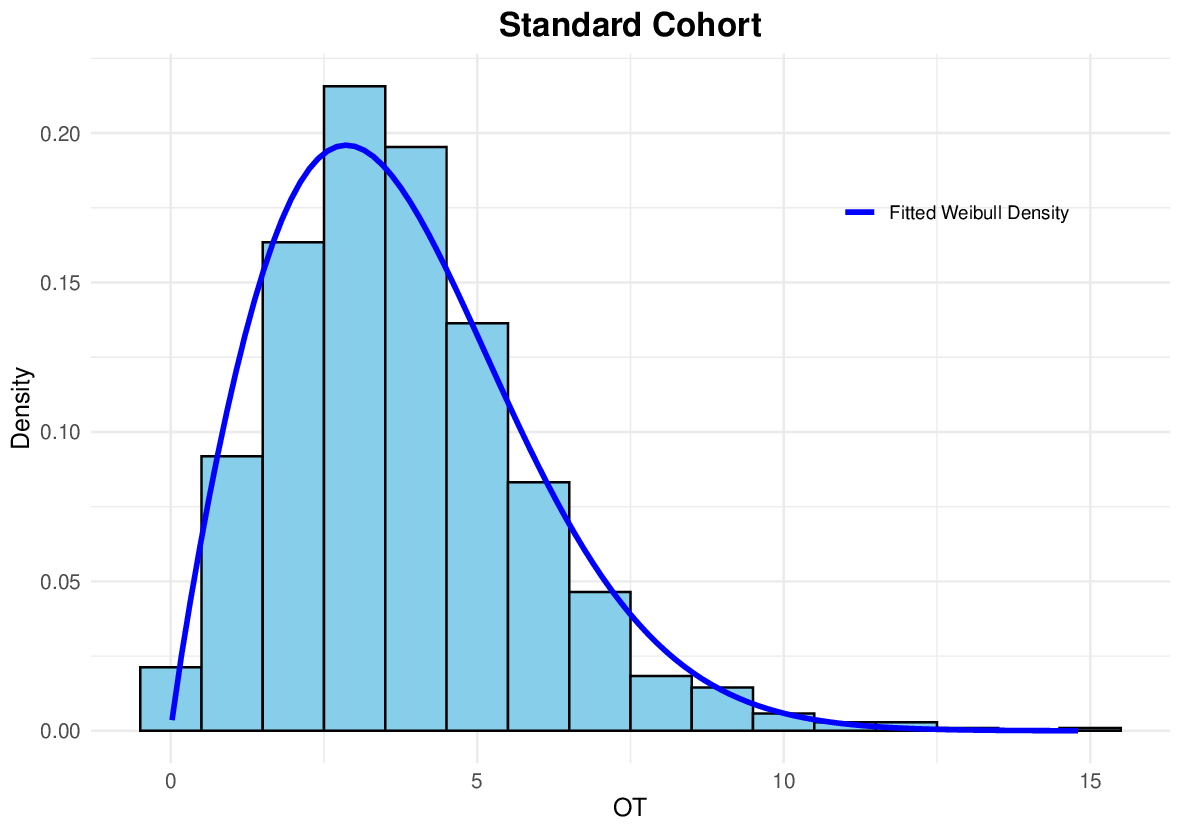}}}
\subfigure[Initial cohort.]{%
\resizebox*{7.5cm}{!}{\includegraphics{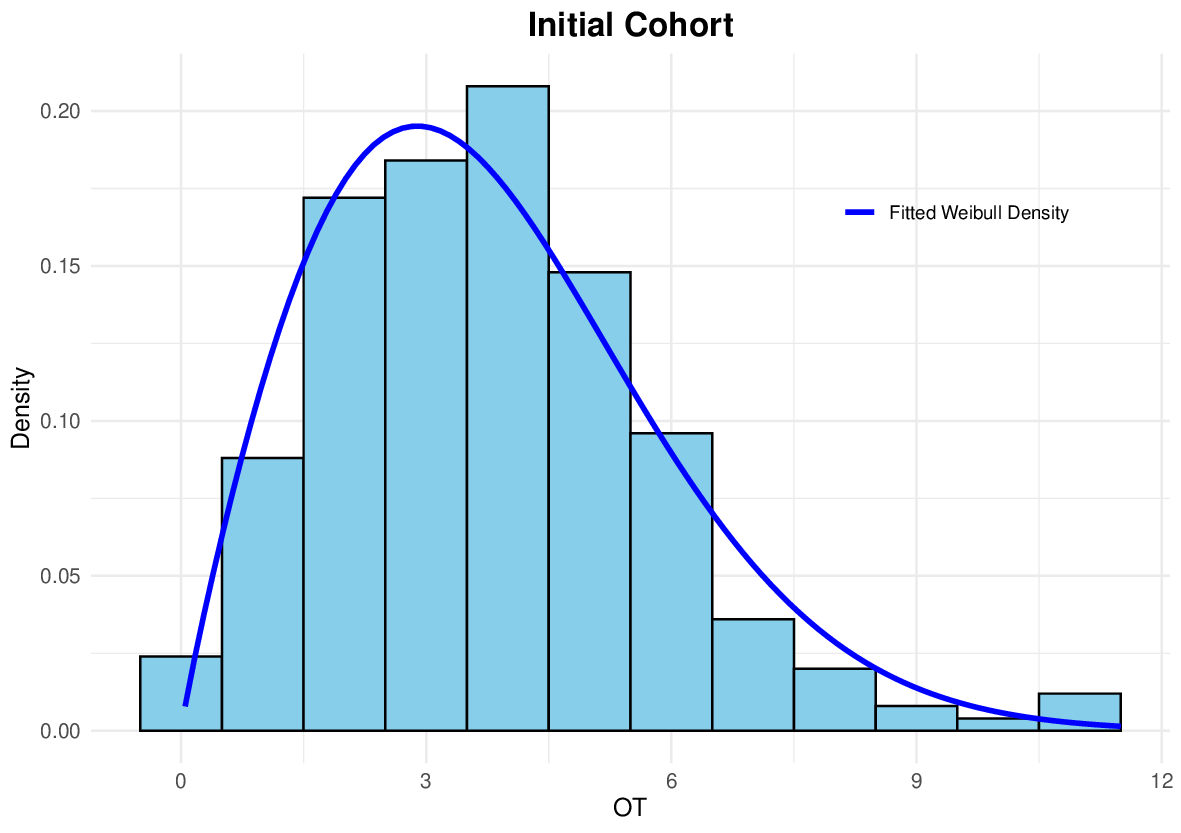}}}
\caption{Checking the Weibull distribution for the operative times.}
\label{fig:GoF}
\end{figure}

We begin by checking the statistical assumptions of the proposed SLCA approach. In particular, we assume that the outcomes of consecutive patients are independent and OT follows a Weibull distribution. The first assumption can be checked through ACF and PACF plots. According to these plots  (not presented here), we observed that the autocorrelation among consecutive patients is negligible and an independent assumption is not violated. To assess the suitability of a Weibull distribution for the operative times, Figure \ref{fig:GoF}(a) and (b) display histograms of the observed OTs with the fitted Weibull density superimposed for each cohort. These figures suggest that the Weibull distribution is a reasonable fit for modelling surgery times in both cohorts. Consequently, the statistical assumptions needed for applying the proposed SLCA technique hold for the underlying colorectal surgery data.

 Table \ref{tab:data} presents the model's parameter estimates, asymptotic standard errors (ASEs), and ACIs for the initial cohort when $\lambda=0.05$. This choice of $\lambda$ is based on the findings from the numerical analysis in Section \ref{sec:compar}. The table indicates that the effect of the risk factor BMI on surgery outcome is statistically significant. Specifically, the negative estimate of $\beta$ implies that higher BMI leads to longer surgery duration. This finding agrees with the scatter plot in Figure \ref{fig:RD}(b).  Given these estimated parameters, the operative time for patients in the initial cohort with a BMI of $12.9$ (the lowest BMI among all patients) is expected to be $4.33$ hours on average, whereas, for those with a BMI of $56.1$ (the highest BMI among all patients), the expected surgery time is $6.25$ hours. This justifies the need for risk adjustment and explains why any comparison that does not account for the risk factor may be misleading. Furthermore, the parameters of the standard cohort are set to be $\gamma_S = 0.1099$, $\eta_S = 1.9220$, and $\beta_S = - 0.0201$.

\begin{table}[]
\centering
\setlength{\tabcolsep}{10pt} 
 \renewcommand{\arraystretch}{1}
\caption{Analysis of colorectal surgery dataset.}
\begin{tabular}{cccccc}
\hline\hline
&    $\gamma_N $ &    & $\eta_N $   &  & $\beta_N$ \\
\hline
Estimate [ASE]&    0.0722  [0.0211]  &   & 1.7859 [0.0881]& &-0.0152 [0.0070]  \\
95\% \text{ACI} &   (0.0309, 0.1135)  &   & (1.6133, 1.9585) & & (-0.0289, -0.0015)   \\
 \hline\hline
\end{tabular}
\label{tab:data}
\end{table}

\begin{figure}[]
 \centering
\subfigure[Mean operative time against time.]{%
\resizebox*{5.8cm}{!}{\includegraphics{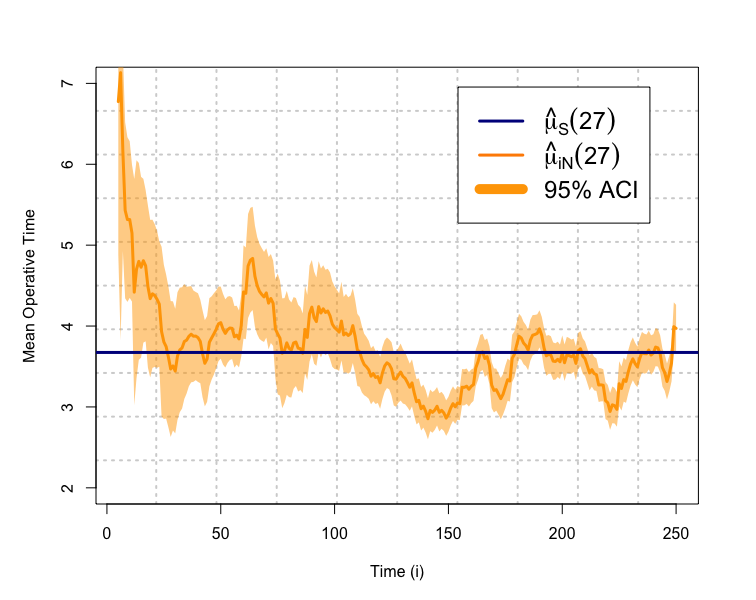}}}
\subfigure[Relative risk against time.]{%
\resizebox*{5.8cm}{!}{\includegraphics{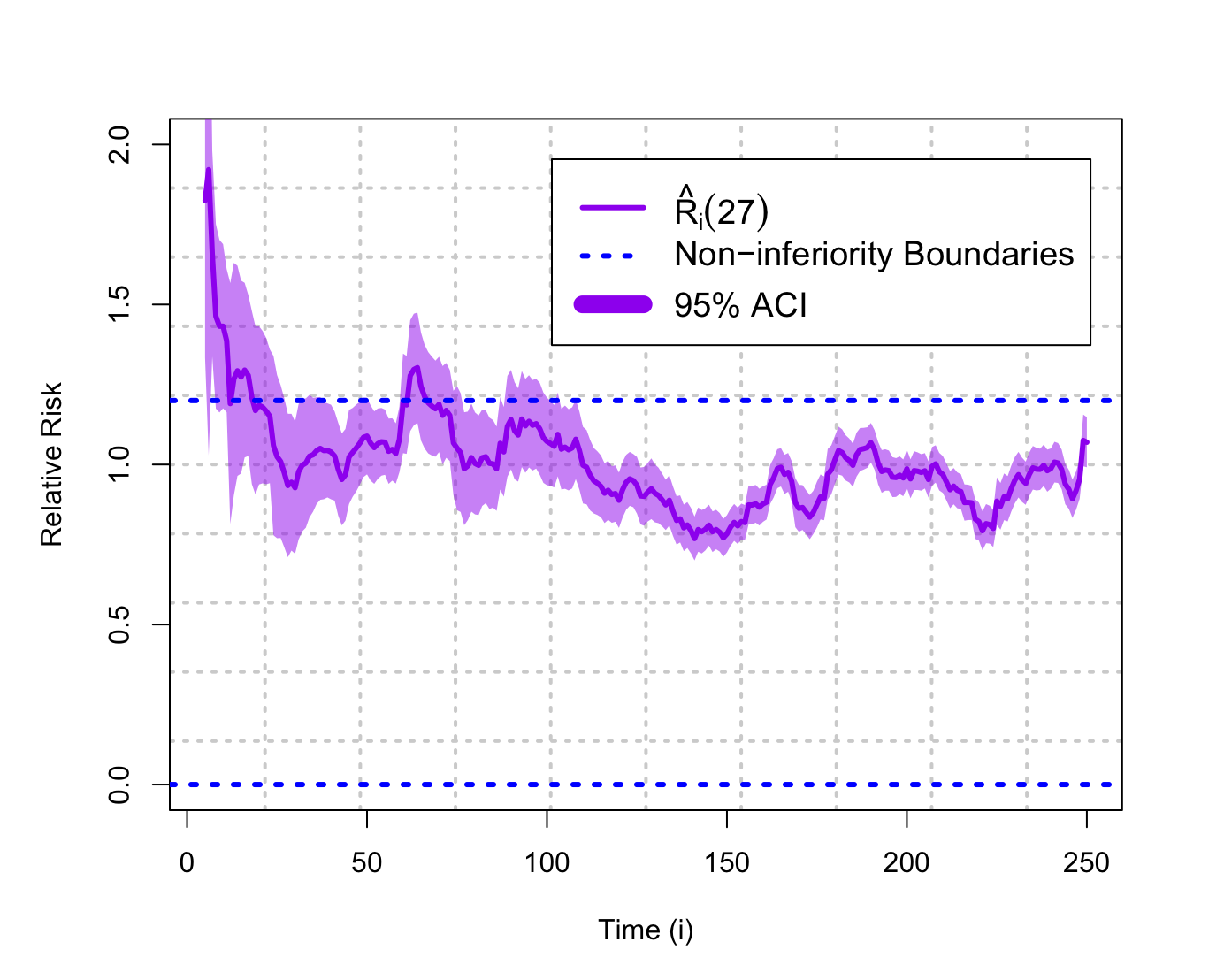}}}
\subfigure[Probability of noninferiority against time.]{%
\resizebox*{5.8cm}{!}{\includegraphics{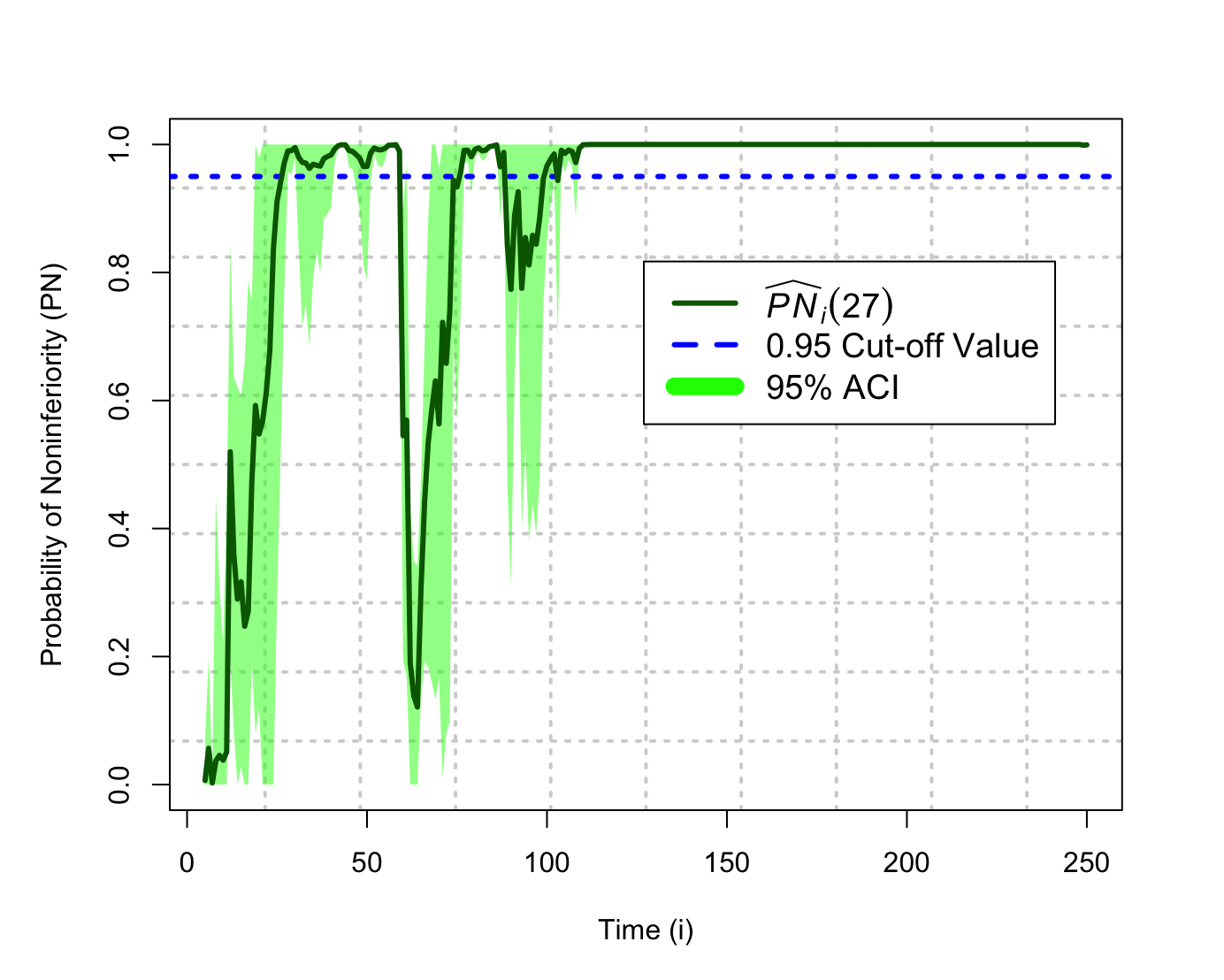}}}
\caption{Analysis of the colorectal surgery dataset.}
\label{fig:SLC}
\end{figure}

In what follows, we apply the proposed SLCA approach to assess the learning process associated with operating on an average patient with a BMI of $x=27$. Then, we used the WEE approach with $\lambda=0.05$, starting from the first surgery and progressively incorporating subsequent outcomes up to the 250-th case. Additionally, we selected the PN (probability of noninferiority) metric with $\epsilon=0.2$ (i.e., $\delta_L=0$ and $\delta_U=1.2$) to analyze the data. This choice implies that learning has occurred if the trainee's performance is within an acceptable margin of 20\% (or better) compared to the standard performance for an average patient. Figures \ref{fig:SLC}(a)-(c) present the estimated RMOT $\hat{\mu}_{iN}(27)$, relative risk $\hat{R}_i(27)$, and PN metric $\widehat{PN}_i(27)$ over time for the cases $i=1, \ldots, 250$. The $95\%$ ACIs are also shown in these plots.

Panel (a) of Figure \ref{fig:SLC} displays the estimated mean operative time for an average patient over time, comparing the new performance (orange) with the established performance (solid blue line). According to this figure, the estimated mean operative time for an average patient under standard performance is $3.71$ hours. Initially, the RMOT for the new performance is significantly higher than for the established performance. However, as case numbers increase, the RMOT decreases, suggesting evidence of a learning process, and gradually approaches the established performance level.  The figure also shows that the confidence intervals around the estimated performance are wider at the beginning, indicating higher uncertainty, and then narrow over time as more data become available. This reflects the nature of the WEE approach, where estimates become more precise with accumulating data. Panel (b) of Figure \ref{fig:SLC} shows the estimated relative risk $\hat{R}_i(27) = \frac{\hat{\mu}_{iN}(27)}{\mu_{S}(27)}$ over time along with the noninferiority boundaries. The relative risk starts above one, and as the case number increases, it decreases and eventually stabilizes within the noninferiority boundaries, with both the relative risk and its confidence intervals fully contained within this region.

Panel (c) of Figure \ref{fig:SLC} illustrates the $\widehat{\text{PN}}$ metric against time, along with a cut-off value of $0.95$. Initially, $\widehat{\text{PN}}$ is low, but, as expected from the behavior of $\hat{R}_i(27)$ in panel (b), $\widehat{\text{PN}}$ gradually rises over time and eventually stabilizes at one. Interestingly, there is a drop in $\widehat{\text{PN}}$ between cases $i=50$ and $i=75$, which aligns with a temporary increase in mean operative time observed in panel (a) during the same period. This suggests that fluctuations in $\widehat{\text{PN}}$ are directly influenced by changes in the mean operative time, emphasizing the sensitivity of $\widehat{\text{PN}}$ to short-term performance changes. Looking retrospectively at panel (c), $\widehat{\text{PN}}_i(27)$ surpasses the $0.95$ cut-off threshold at case $i=100$ and remains above this level for the rest of the cases. This time point can be referred to as the "expertise time", indicating that the new performance has gained sufficient experience to operate on patients with a BMI of $x=27$ with results that are noninferior to the standard performance, which coincides with the assessment of the raw data presented in Figure \ref{fig:RD}(c). It is finally important to note that according to this figure, there are times before $i=100$ that it looks like learning has happened, but then we see some deterioration later. This phenomenon underscores the importance of continued monitoring when applying the method prospectively. Rather than stopping the monitoring immediately after detecting a positive signal, it is crucial to maintain surveillance for a sufficient period to ensure that performance is sustained.

We conclude by emphasizing that the proposed SLCA approach is capable of providing comprehensive information about the learning process. For example, beyond the insights provided by the analysis above, one might wish to assess the stabilized performance of the trainee (at time $i=100$) across patients with different conditions (e.g., a range of BMIs). The proposed method is well-equipped to handle such assessments as well. Thus, not only does it offer insights into the effect of time on the trainee's performance, but it also provides clinicians with valuable information about performance across a range of patient characteristics through the estimation of various performance measures. These measures are easy to understand and interpret.  In cases where agreement or noninferiority is not established, the approach offers helpful diagnostic tools like RMOT and relative risk plots. These tools not only provide significant insights into performance evaluation but also serve as valuable diagnostics within the proposed SLCA framework. Unlike the LC-CUSUM approach, which lacks this level of detail, the proposed SLCA approach delivers a more granular view of the learning process and overall performance.

\section{Comparison study}
\label{sec:compar}
 
This subsection compares the performance of the proposed methodology with a risk-adjusted LC-CUSUM approach. As mentioned earlier in the introduction, one of the most widely used SLC techniques is the LC-CUSUM approach. However, in its original definition, the LC-CUSUM is not a risk-adjusted approach for continuous outcomes. We will begin by reviewing the original definition of LC-CUSUM and then propose an adaptation to account for the effect of risk factors. This will ensure that the comparison with our proposed method is fair.

Assume that the outcome of interest, $Y$, follows a normal distribution with mean $\mu$ and standard deviation $\sigma$, where the surgeon's performance is measured by $\mu$. The LC-CUSUM approach assumes three levels of performance: adequate performance ($\mu = \mu_0$), inadequate performance ($\mu = \mu_1$), and a performance deviating acceptably by $\epsilon^{'}$ from the adequate level, i.e., $\mu = \mu_0 + \epsilon^{'}$. Without loss of generality, we assume that $\mu_1 > \mu_0$. Similar to other monitoring approaches, LC-CUSUM tests a hypothesis at each data point. The hypotheses are defined as $H_0: \mu = \mu_0 + \epsilon^{'}$ versus $H_1: \mu = \mu_0$. This order of hypotheses stems from the fact that trainees are more likely to start with inadequate performance and improve with experience. Biau and Porcher \cite{biau2010method} suggest calculating the following quantity at each time point
\begin{align}\label{equ:wt}
v_i = \frac{y_i - (\mu_0 + \frac{\epsilon^{'}}{2})}{\sigma},
\end{align}
where $v_i$ can be viewed as the standardized residual between the actual outcome at time $i$ and the expected outcome under adequate performance, deviated by a magnitude of $\frac{\epsilon^{'}}{2}$. Consistent with control chart literature, they assume that the deviation size is proportional to the standard deviation $\sigma$, i.e., $\epsilon^{'} = g \sigma$ for $g > 0$. Using this, $v_i$ in \eqref{equ:wt} can be rewritten as
\begin{align*} 
v_i = \frac{y_i - \mu_0}{\sigma} - \frac{g}{2}.
\end{align*}

They then propose using the following recursive statistic
\begin{align*} 
s_i = \min\{0, s_{i-1} + v_i\}, \quad \text{where} \quad s_0 = 0, 
\end{align*}
as the monitoring statistic. Given a predefined cut-off value $h$, the LC-CUSUM triggers an alarm as soon as $|s_i| > h$. Such a signal suggests sufficient evidence in the data to support the alternative hypothesis, indicating that the surgeon has become proficient. The time index $i$ at which the signal occurs can be referred to as the point of becoming an expert.

In what follows, we adapt the LC-CUSUM approach based on the setting considered in this paper. Given the assumptions of the proposed SLCA approach, we know that the conditional distribution of $Y$ given $\textbf{x}$ is a Weibull distribution with mean $\mu(\textbf{x})$, as given in \eqref{equ:meanweibull}. The corresponding standard deviation is given by
\begin{align*} 
\text{SD}(Y|\textbf{x}) = \sqrt{\frac{\Gamma\left(\frac{2}{\eta}+1\right)}{\theta^{2 / \eta}} - \mu^2(\textbf{x})}.
\end{align*}

In this context, the actual performance is represented by $\mu_N(\textbf{x})$ (analogous to $\mu$ in the LC-CUSUM definition), and adequate performance is defined as the expected standard performance $\mu_S(\textbf{x})$ (analogous to $\mu_0$ in the LC-CUSUM definition). Inadequate performance is modeled as being proportional to $\mu_S(\textbf{x})$, e.g., $2 \times \mu_S(\textbf{x})$ (analogous to $\mu_1$ in the LC-CUSUM definition). The performance deviated by $\epsilon$ is given by $\mu_S(\textbf{x})(1 + \epsilon)$ (analogous to $\mu_0+\epsilon^{'}$ in the LC-CUSUM definition). Thus, the null hypothesis would be $H_0: \mu_N(\textbf{x}) = \mu_S(\textbf{x})(1 + \epsilon)$ and the alternative hypothesis would be $H_1: \mu_N(\textbf{x}) = \mu_S(\textbf{x})$. In this framework, the deviation magnitude $\epsilon'$ in the LC-CUSUM approach, which was originally proportional to the standard deviation of the outcome (i.e., $\epsilon' = g \sigma$), is now replaced by $\epsilon \times \mu_S(\textbf{x})$, which is proportional to the expected outcome. Based on these definitions, we suggest using the following monitoring statistic
\begin{align*} 
s_i(\textbf{x}) = \min\{0, s_{i-1}(\textbf{x}) + v_i(\textbf{x})\} \qquad \text{where} \qquad v_i(\textbf{x}) = \frac{y_i - \widehat{\mu}_S(\textbf{x})(1+\frac{\epsilon}{2})}{\widehat{SD}(Y_S|\textbf{x})}, \qquad \text{for} \qquad i=1,2,\ldots.
\end{align*}

The adapted LC-CUSUM approach signals when $|s_i(\textbf{x})| > h(\textbf{x})$, where the cut-off value $h(\textbf{x})$ should be specified to ensure that certain performance measures meet their targets.

In this study, we use two performance measures that are well-suited for healthcare quality monitoring. These measures are the probability of a false alarm (PFA), representing the signal probability under an inadequate performance scenario, and the probability of successful detection (PSD), representing the signal probability under a learning process scenario, both defined over a specified time frame $[1, t]$ (see \cite{marshall2004statistical}). More precisely,  PFA$_t$ represents the probability that a false alarm (an alarm triggered when performance is inadequate) will occur by time $t$. On the other hand, PSD$_t$ is the probability that the method will correctly signal a true alarm (an alarm when performance is adequate) within a specific time period after the process has shifted from an inadequate state to an adequate one, assuming no false alarms occurred during the inadequate performance period. We use these measures to compare our approach with the adapted LC-CUSUM method, as Biau and Porcher \cite{biau2010method} also did. These metrics can also help in determining an appropriate decision cut-off value for both methods to make a fair comparison. To do this, we conducted a simulation study.

\begin{figure}[]
    \centering
    \includegraphics[scale=0.6]{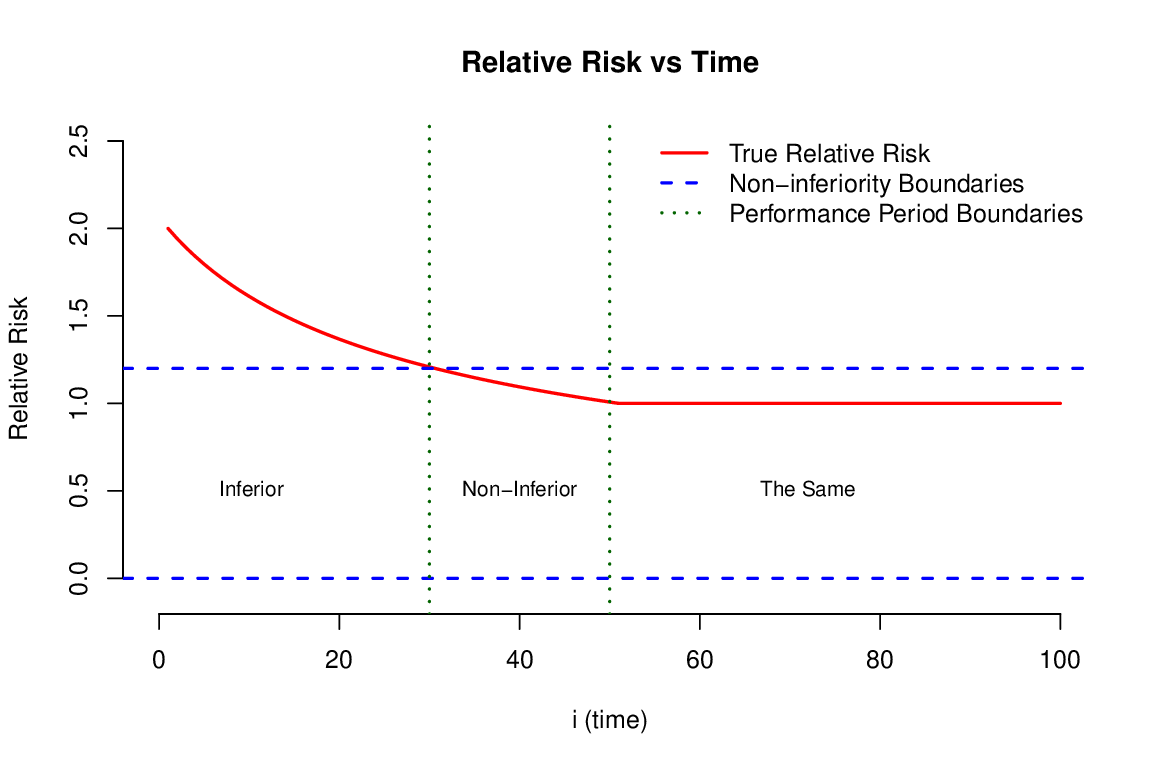}  
    \caption{The true $R_i(27)$ used in simulation study under the learning process scenario along with the noninferiority boundaries.}
    \label{fig:rsim}  
\end{figure}

 \begin{table}[]
\centering
\setlength{\tabcolsep}{15pt} 
\renewcommand{\arraystretch}{1}
\caption{Comparison study results based on probability of false alarm under the inadequate performance and the probability of successful detection under the learning process.}
\begin{tabular}{lcccccc}
\hline\hline
 & \multicolumn{6}{c}{$x=23$} \\ 
 \cline{2-7}
 &  & Inadequate performance scenario  & & \multicolumn{3}{c}{Learning process scenario}  \\
 \cline{3-3}\cline{5-7}
& $h(x)$ & PFA$_{100}$ & & PSD$_{20}$ & PSD$_{50}$ & PSD$_{70}$     \\
 \cline{3-3}\cline{5-7}
LC-CUSUM& 3.80 & 0.039 & & 0.378 & 0.775 & 0.879 \\
SLCA with $\lambda=0.01$ & 0.78 & 0.043 & & 0.045  & 0.359 & 0.923   \\
SLCA with $\lambda=0.05$& 0.75 &  0.047& & 0.336 & 0.841& 0.899  \\
SLCA with $\lambda=0.10$ & 0.75 &0.067 & & 0.593& 0.745& 0.746 \\
\hline
& \multicolumn{6}{c}{$x=27$} \\
 \cline{3-7}
 & $h(x)$ & PFA$_{100}$ & & PSD$_{20}$ & PSD$_{50}$ & PSD$_{70}$\\
 LC-CUSUM& 4.00 &  0.049 & & 0.454 & 0.798 &  0.839  \\
SLCA with $\lambda=0.01$& 0.75 &  0.030 & & 0.037   & 0.444  & 0.957   \\
SLCA with $\lambda=0.05$& 0.75 & 0.030  & & 0.382  &0.921  & 0.933  \\
SLCA with $\lambda=0.10$&0.75  &0.030  &  & 0.684  & 0.822 & 0.823  \\
\hline
& \multicolumn{6}{c}{$x=37.5$} \\
 \cline{3-7}
& $h(x)$ &PFA$_{100}$ & & PSD$_{20}$ & PSD$_{50}$ & PSD$_{70}$\\
 LC-CUSUM& 6.30 &  0.043 &  & 0.590 & 0.636 & 0.636 \\
SLCA with $\lambda=0.01$& 0.75 &  0.052 & & 0.081  & 0.343  & 0.876  \\
SLCA with $\lambda=0.05$& 0.75 & 0.058 & & 0.301  & 0.754 & 0.816   \\
SLCA with $\lambda=0.10$&0.95 & 0.040 & & 0.431 & 0.849 & 0.882 \\
 \hline\hline
\end{tabular}
\label{tab:sim}
\end{table}

We designed the simulation study based on the colorectal surgery data discussed in Section \ref{sec:example}. In this study, the risk factor $x$ (BMI) takes values from the set $\{13, \ldots, 56\}$. We applied the PN metric with $\epsilon=0.2$ to compare the trainee surgeon with a standard performance and assess the learning process. For this simulation, we set $t=100$ surgeries and assumed the following parameters: $\gamma_S=0.2$, $\eta_S = \eta_N = 2$, and $\beta_S = \beta_N = -0.05$. For the inadequate performance scenario, when calculating PFA, we set $\gamma_N=0.05$. Under this setup, the model yielded $R_i(x)=2$ for $i=1, 2, \ldots, 100$. On the other hand, to simulate the learning process scenario, we assumed learning occurs during the first half of the surgeries, after which the trainee surgeon's performance matches the standard. Specifically, for $i = 1, 2, \ldots, 50$, the parameter $\gamma_N$ starts at $0.05$ and increases incrementally by $0.003$ until $t=51$, where $\gamma_N$ reaches $0.2$. For the remaining surgeries ($i=51, 52, \ldots, 100$), we set $\gamma_N = \gamma_S = 0.2$. To calculate PN, we selected three BMI values: $x \in \{23, 27, 37.5\}$ representing the 25-th, 50-th, and 90-th percentiles of the BMI distribution in the colorectal case study. We also used three values for the smoothing parameter: $\lambda \in \{0.01,0.05,0.1\}$. Figure \ref{fig:rsim} illustrates the effect of these settings within the learning process scenario on the relative risk at $x=27$ over time, along with the noninferiority boundaries. According to the figure, the relative risk enters the noninferiority interval at $i=31$, indicating a shift from inadequate to a clinically noninferior performance, and remains within this interval until the end of the study ($t=100$). At $i=51$, the relative risk reaches a value of one, meaning the trainee’s performance matches the standard exactly.

The results of the comparison study are presented in Table \ref{tab:sim}. In this table, PFA$_{100}$ shows the observed false signal rate by time $t=100$ when performance is inadequate throughout the period $[1,100]$. Additionally, PSD$_{20}$ represents the estimated probability of successfully detecting adequate performance within 20 surgeries after $i=30$, the point where the performance shifts from inferior to noninferior. This refers to signalling within the interval $[31,50]$, assuming no false signals were observed in the interval $[1,30]$. Similarly, PSD$_{50}$ (PSD$_{70}$) represents the estimated probability of successfully detecting noninferior performance by case number 50 (or 70) after $i=30$, i.e., signalling within the intervals $[31,80]$ ($[31,100]$), assuming no false signals were observed up to and including $i=30$. These probabilities are obtained based on 10,000 simulation runs. The table also includes the cut-off values $h(x)$, which were set to achieve a PFA$_{100}$ around $0.05$, specifically within the range PFA$_{100} \in [0.03, 0.07]$.

According to the table, increasing the smoothing parameter $\lambda$ in our method significantly impacts the performance metrics. For instance, when $\lambda$ is set to 0.01, PSD values for early detection intervals (PSD$_{20}$ and PSD$_{50}$) are relatively low, indicating that our method struggles to quickly detect performance improvements. As $\lambda$ increases to 0.05, PSD values generally improve across all time intervals, reflecting stronger detection ability. Further increasing $\lambda$ to 0.1 enhances sensitivity to performance changes in some cases, particularly early on, but can reduce detection ability over longer intervals. These findings align with the nature of the WEE approach used. Overall, $\lambda=0.05$ appears to offer the best balance that provides relatively high PSD values across time intervals. 

Another interesting trend in the table is the effect of the risk factor $x$ on the performance measures. As $x$ gets closer to $27$, the detection ability of the proposed SLCA approach improves. This is likely due to the amount of information available in the dataset at these risk factor values. Since $x=27$ represents the median, there is more data available for patients with characteristics close to this value than for those at the 25th percentile ($x=23$) or the 90th percentile ($x=37.5$). This greater amount of data enhances the ability to detect performance changes effectively.

With $\lambda=0.05$, while the LC-CUSUM approach performs slightly better than the SLCA approach in the early phase of the learning process, the SLCA approach generally outperforms LC-CUSUM as the learning process progresses across different risk factor values. For example, at $x=27$, there is a 45\% chance that the LC-CUSUM method will correctly signal an alarm within the first 20 observations after the change point, compared to a 38\% chance with the SLCA approach. However, the SLCA approach shows improved detection rates over longer intervals, with a 92\% probability of detection within 50 observations and a 93\% probability within 70 observations after the change point. In contrast, LC-CUSUM achieves these with probabilities of 79\% and 83\%, respectively, indicating that the proposed method becomes more effective in accurately detecting changes over time.

 \section{Conclusion}
Comparative probability metrics (CPM) have been increasingly developed and used to compare the characteristics of different populations. We propose the use of CPM within a learning curve framework to evaluate the performance of a trainee relative to a standard benchmark. We used operative time as the outcome of interest and employed a Weibull regression to account for patient-level characteristics. The proposed method sequentially estimates performance measures over time to track the trainee's progress and identify when they reach sufficient expertise. The method also includes visualizations that plot performance measures against time and patient characteristics, providing valuable insights for clinicians to better understand the learning process and performance trends. The proposed method can assess learning curves retrospectively using complete datasets or prospectively by applying it as new outcomes become available, assuming a known standard performance model at the start. In addition, in cases where the standard performance needs to be estimated, the proposed method can be readily extended to incorporate the uncertainty introduced by its estimation into the learning curve assessment. The effectiveness of the proposed method was demonstrated through a case study concerning colorectal surgeries as well as a numerical study. 

Although the methodology is broadly applicable, there are several potential extensions to explore in future work. The proposed method was illustrated assuming the outcome of interest is operative time, but it may be used for any right skewed outcome well-characterized by the Weibull distribution. However, in certain surgeries, the outcome of interest may be continuous but not right-skewed or even binary, categorical, or ordinal. An immediate direction for further research is to extend the proposed method to handle these different data types.

\section*{Conflicts of Interest}
The authors declare no conflicts of interest.

\section*{Data Availability Statement}

The dataset analyzed in this paper is publicly available and can also be obtained from the corresponding author upon request.

\bibliography{wileyNJD-AMA}

\begin{thebibliography}{10}
\providecommand \doibase [0]{http://dx.doi.org/}%

\bibitem{wright1936factors}
Wright TP. Factors affecting the cost of airplanes. {\it Journal of the
  aeronautical sciences.} 1936\string;3(4)\string:122--128.

\bibitem{harrysson2014systematic}
Harrysson IJ, Cook J, Sirimanna P, Feldman LS, Darzi A, Aggarwal R. Systematic
  review of learning curves for minimally invasive abdominal surgery: a review
  of the methodology of data collection, depiction of outcomes, and statistical
  analysis. {\it Annals of surgery.} 2014\string;260(1)\string:37--45.

\bibitem{valsamis2018learning}
Valsamis EM, Chouari T, O’Dowd-Booth C, Rogers B, Ricketts D. Learning curves
  in surgery: variables, analysis and applications. {\it Postgraduate medical
  journal.} 2018\string;94(1115)\string:525--530.

\bibitem{ahn2021learning}
Ahn Y, Lee S, Son S, Kim H. Learning curve for interlaminar endoscopic lumbar
  discectomy: a systematic review. {\it World Neurosurgery.}
  2021\string;150\string:93--100.

\bibitem{burghgraef2022learning}
Burghgraef TA, Sikkenk DJ, Verheijen PM, Moumni ME, Hompes R, Consten EC. The
  learning curve of laparoscopic, robot-assisted and transanal total mesorectal
  excisions: a systematic review. {\it Surgical Endoscopy.}
  2022\string;36(9)\string:6337--6360.

\bibitem{patel2023scoping}
Patel YS, Mistry N, Farrokhyar F, Simunovic M, Hanna WC. Scoping review of
  learning curve methods in minimally invasive thoracic surgery. {\it Global
  Surgical Education-Journal of the Association for Surgical Education.}
  2023\string;2(1)\string:81.

\bibitem{woodall2015monitoring}
Woodall WH, Fogel SL, Steiner SH. The monitoring and improvement of
  surgical-outcome quality. {\it Journal of Quality Technology.}
  2015\string;47(4)\string:383--399.

\bibitem{lim2002assessing}
Lim T, Soraya A, Ding L, Morad Z. Assessing doctors’ competence: application
  of CUSUM technique in monitoring doctors’ performance. {\it International
  Journal for Quality in Health Care.} 2002\string;14(3)\string:251--258.

\bibitem{knoth2019risk}
Knoth S, Wittenberg P, Gan FF. Risk-adjusted CUSUM charts under model error.
  {\it Statistics in medicine.} 2019\string;38(12)\string:2206--2218.

\bibitem{gomon2024inspecting}
Gomon D, Sijmons J, Putter H, et al. Inspecting the quality of care: a
  comparison of CUSUM methods for inter hospital performance. {\it Health
  Services and Outcomes Research Methodology.}
  2024\string;24(3)\string:281--303.

\bibitem{woodall2021overview}
Woodall WH, Rakovich G, Steiner SH. An overview and critique of the use of
  cumulative sum methods with surgical learning curve data. {\it Statistics in
  medicine.} 2021\string;40(6)\string:1400--1413.

\bibitem{forbes2007learning}
Forbes TL, Chu MW, Lawlor DK, DeRose G, Harris KA. Learning curve analysis of
  thoracic endovascular aortic repair in relation to credentialing guidelines.
  {\it Journal of vascular surgery.} 2007\string;46(2)\string:218--222.

\bibitem{forbes2004cumulative}
Forbes TL, DeRose G, Kribs SW, Harris KA. Cumulative sum failure analysis of
  the learning curve with endovascular abdominal aortic aneurysm repair. {\it
  Journal of vascular surgery.} 2004\string;39(1)\string:102--108.

\bibitem{forbes2007association}
Forbes TL, DeRose G, Lawlor DK, Harris KA. The association between a
  surgeon’s learning curve with endovascular aortic aneurysm repair and
  previous institutional experience. {\it Vascular and endovascular surgery.}
  2007\string;41(1)\string:14--18.

\bibitem{biau2008quantitative}
Biau D, Williams S, Schlup M, Nizard R, Porcher R. Quantitative and
  individualized assessment of the learning curve using LC-CUSUM. {\it Journal
  of British Surgery.} 2008\string;95(7)\string:925--929.

\bibitem{biau2010method}
Biau DJ, Porcher R. A method for monitoring a process from an out of control to
  an in control state: Application to the learning curve. {\it Statistics in
  medicine.} 2010\string;29(18)\string:1900--1909.

\bibitem{ward2014analysis}
Ward ST, Mohammed MA, Walt R, Valori R, Ismail T, Dunckley P. An analysis of
  the learning curve to achieve competency at colonoscopy using the JETS
  database. {\it Gut.} 2014\string;63(11)\string:1746--1754.

\bibitem{dong2021comprehensive}
Dong Z, Sun H, Li B, et al. Comprehensive evaluation of the learning curve to
  achieve satisfactory adenoma detection rate. {\it Journal of gastroenterology
  and hepatology.} 2021\string;36(6)\string:1649--1655.

\bibitem{eisenberg2017applying}
Eisenberg VH, Alcazar JL, Arbib N, et al. Applying a statistical method in
  transvaginal ultrasound training: lessons from the learning curve cumulative
  summation test (LC-CUSUM) for endometriosis mapping. {\it Gynecological
  Surgery.} 2017\string;14\string:1--10.

\bibitem{govindarajulu2018survival}
Govindarajulu U, Bedi S, Kluger A, Resnic F. Survival analysis of hierarchical
  learning curves in assessment of cardiac device and procedural safety. {\it
  Statistics in Medicine.} 2018\string;37(28)\string:4185--4199.

\bibitem{wu2022analyzing}
Wu CJ, Huang KJ, Chang WC, Li YX, Wei LH, Sheu BC. Analyzing the learning curve
  of vaginal pelvic reconstruction surgery with and without mesh by the
  cumulative summation test (CUSUM). {\it Scientific Reports.}
  2022\string;12(1)\string:7025.

\bibitem{turan2015season}
Turan O, Babazade R, Eshraghi Y, You J, Turan A, Remzi F. Season and vitamin D
  status do not affect probability for surgical site infection after colorectal
  surgery. {\it European Surgery.} 2015\string;47\string:341--345.

\bibitem{jimenez2013learning}
Jim{\'e}nez-Rodr{\'\i}guez RM, D{\'\i}az-Pav{\'o}n JM, Juan d.~l. P.~dF,
  Prendes-Sillero E, Dussort HC, Padillo J. Learning curve for robotic-assisted
  laparoscopic rectal cancer surgery. {\it International journal of colorectal
  disease.} 2013\string;28\string:815--821.

\bibitem{gezer2016cusum}
Gezer S, Avc{\i} A, T{\"u}rktan M. Cusum analysis for learning curve of
  videothoracoscopic lobectomy. {\it Open Medicine.}
  2016\string;11(1)\string:574--577.

\bibitem{van2016outcome}
Poel v.~dMJ, Besselink MG, Cipriani F, et al. Outcome and learning curve in 159
  consecutive patients undergoing total laparoscopic hemihepatectomy. {\it JAMA
  surgery.} 2016\string;151(10)\string:923--928.

\bibitem{wang2016learning}
Wang M, Meng L, Cai Y, et al. Learning curve for laparoscopic
  pancreaticoduodenectomy: a CUSUM analysis. {\it Journal of gastrointestinal
  surgery.} 2016\string;20(5)\string:924--935.

\bibitem{kim2021evaluation}
Kim S, Yoon YS, Han HS, Cho JY, Choi Y, Lee B. Evaluation of a single
  surgeon’s learning curve of laparoscopic pancreaticoduodenectomy:
  risk-adjusted cumulative summation analysis. {\it Surgical endoscopy.}
  2021\string;35\string:2870--2878.

\bibitem{dimitrovska2022learning}
Dimitrovska NT, Bao F, Yuan P, Hu S, Chu X, Li W. Learning curve for two-port
  video-assisted thoracoscopic surgery lung segmentectomy. {\it Interactive
  CardioVascular and Thoracic Surgery.} 2022\string;34(3)\string:402--407.

\bibitem{lin2023cusum}
Lin PL, Zheng F, Shin M, Liu X, Oh D, D’Attilio D. CUSUM learning curves:
  what they can and can’t tell us. {\it Surgical Endoscopy.}
  2023\string;37(10)\string:7991--7999.

\bibitem{rakovich2022comment}
Rakovich G, Woodall WH, Steiner S. Comment on the CUSUM surgical learning curve
  analysis in Dimitrovska et al.(2022). {\it Interactive CardioVascular and
  Thoracic Surgery.} 2022\string;35(2)\string:ivac184.

\bibitem{steiner2014risk}
Steiner SH. Risk-adjusted monitoring of outcomes in health care. {\it
  Statistics in action: A Canadian outlook.} 2014\string;14\string:225--41.

\bibitem{steiner2000monitoring}
Steiner SH, Cook RJ, Farewell VT, Treasure T. Monitoring surgical performance
  using risk-adjusted cumulative sum charts. {\it Biostatistics.}
  2000\string;1(4)\string:441--452.

\bibitem{steiner2001risk}
Steiner SH, Cook RJ, Farewell VT. Risk-adjusted monitoring of binary surgical
  outcomes. {\it Medical Decision Making.} 2001\string;21(3)\string:163--169.

\bibitem{khan2014measuring}
Khan N, Abboudi H, Khan MS, Dasgupta P, Ahmed K. Measuring the surgical
  ‘learning curve’: methods, variables and competency. {\it BJU
  international.} 2014\string;113(3)\string:504--508.

\bibitem{inaba2019operative}
Inaba CS, Koh CY, Sujatha-Bhaskar S, Gallagher S, Chen Y, Nguyen NT. Operative
  time as a marker of quality in bariatric surgery. {\it Surgery for Obesity
  and Related Diseases.} 2019\string;15(7)\string:1113--1120.

\bibitem{uecker2013comparable}
Uecker J, Luftman K, Ali S, Brown C. Comparable operative times with and
  without surgery resident participation. {\it Journal of Surgical Education.}
  2013\string;70(6)\string:696--699.

\bibitem{jackson2011does}
Jackson TD, Wannares JJ, Lancaster RT, Rattner DW, Hutter MM. Does speed
  matter? The impact of operative time on outcome in laparoscopic surgery. {\it
  Surgical endoscopy.} 2011\string;25\string:2288--2295.

\bibitem{kim2014operative}
Kim BD, Hsu WK, De~Oliveira~Jr GS, Saha S, Kim JY. Operative duration as an
  independent risk factor for postoperative complications in single-level
  lumbar fusion: an analysis of 4588 surgical cases. {\it Spine.}
  2014\string;39(6)\string:510--520.

\bibitem{lane2020correlation}
Lane J, Whitehurst L, Hameed BZ, Tokas T, Somani BK. Correlation of operative
  time with outcomes of ureteroscopy and stone treatment: a systematic review
  of literature. {\it Current urology reports.} 2020\string;21\string:1--9.

\bibitem{offodile2017impact}
Offodile AC, Aherrera A, Wenger J, Rajab TK, Guo L. Impact of increasing
  operative time on the incidence of early failure and complications following
  free tissue transfer? A risk factor analysis of 2,008 patients from the
  ACS-NSQIP database. {\it Microsurgery.} 2017\string;37(1)\string:12--20.

\bibitem{maggino2018impact}
Maggino L, Liu JB, Ecker BL, Pitt HA, Vollmer~Jr CM. Impact of operative time
  on outcomes after pancreatic resection: a risk-adjusted analysis using the
  American College of Surgeons NSQIP Database. {\it Journal of the American
  College of Surgeons.} 2018\string;226(5)\string:844--857.

\bibitem{joustra2013can}
Joustra P, Meester R, Ophem vH. Can statisticians beat surgeons at the planning
  of operations?. {\it Empirical Economics.} 2013\string;44\string:1697--1718.

\bibitem{balasooriya1994selecting}
Balasooriya U, Abeysinghe T. Selecting between gamma and Weibull
  distributions—an approach based on predictions of order statistics. {\it
  Journal of Applied Statistics.} 1994\string;21(3)\string:17--27.

\bibitem{carroll2003use}
Carroll KJ. On the use and utility of the Weibull model in the analysis of
  survival data. {\it Controlled clinical trials.}
  2003\string;24(6)\string:682--701.

\bibitem{ying2008weibull}
Ying Gs, Heitjan DF. Weibull prediction of event times in clinical trials. {\it
  Pharmaceutical Statistics: The Journal of Applied Statistics in the
  Pharmaceutical Industry.} 2008\string;7(2)\string:107--120.

\bibitem{zhang2012modeling}
Zhang X, Long Q. Modeling and prediction of subject accrual and event times in
  clinical trials: a systematic review. {\it Clinical Trials.}
  2012\string;9(6)\string:681--688.

\bibitem{glance2018variability}
Glance LG, Dutton RP, Feng C, Li Y, Lustik SJ, Dick AW. Variability in case
  durations for common surgical procedures. {\it Anesthesia \& Analgesia.}
  2018\string;126(6)\string:2017--2024.

\bibitem{stevens2020bayesian}
Stevens NT, Lu L, Anderson-Cook CM, Rigdon SE. Bayesian probability of
  agreement for comparing survival or reliability functions with parametric
  lifetime regression models. {\it Quality Engineering.}
  2020\string;32(3)\string:312--332.

\bibitem{stevens2017assessing}
Stevens NT, Steiner SH, MacKay RJ. Assessing agreement between two measurement
  systems: An alternative to the limits of agreement approach. {\it Statistical
  methods in medical research.} 2017\string;26(6)\string:2487--2504.

\bibitem{steiner2014monitoring}
Steiner SH, Mackay RJ. Monitoring risk-adjusted medical outcomes allowing for
  changes over time. {\it Biostatistics.} 2014\string;15(4)\string:665--676.

\bibitem{white1982maximum}
White H. Maximum likelihood estimation of misspecified models. {\it
  Econometrica: Journal of the econometric society.} 1982\string:1--25.

\bibitem{wang2004asymptotic}
Wang X, Eeden vC, Zidek JV. Asymptotic properties of maximum weighted
  likelihood estimators. {\it Journal of Statistical Planning and Inference.}
  2004\string;119(1)\string:37--54.

\bibitem{stevens2020comparing}
Stevens NT, Lu L. Comparing Kaplan-Meier curves with the probability of
  agreement. {\it Statistics in Medicine.}
  2020\string;39(30)\string:4621--4635.

\bibitem{marshall2004statistical}
Marshall C, Best N, Bottle A, Aylin P. Statistical issues in the prospective
  monitoring of health outcomes across multiple units. {\it Journal of the
  Royal Statistical Society Series A: Statistics in Society.}
  2004\string;167(3)\string:541--559.

\end{thebibliography}

\appendix

\bmsection{The closed form of matrix $Q\left(\gamma, \eta, \pmb{\beta} \mid \mathbf{y}, \mathbf{x}, \mathbf{w}\right)$.}
 After some mathematical simplifications, the partial derivatives in \eqref{WEEequ1} can be derived as

\begin{align*} 
\frac{\partial l_{i}}{\partial \gamma}=\frac{1}{\gamma} -  y_{i}^{\eta} e^{\pmb{\beta}^\top \textbf{x}_{i}},\, \frac{\partial l_{i}}{\partial \eta}&=\frac{1}{\eta}+\ln(y_{i})\Big(1-  \gamma  y_{i}^{\eta} e^{\pmb{\beta}^\top \textbf{x}_{i}}\Big),\, \frac{\partial l_{i}}{\partial \beta_{j}}= x_{ij} \Big(1- \gamma y_{i}^{\eta} e^{\pmb{\beta}^\top \textbf{x}_{i}}\Big) \, \text{for $j=1,\ldots,d$}.
\end{align*}

Accordingly, we could rewrite the score functions in \eqref{WEEequ1} and build the following simplified system of equations. It is
\begin{align*} 
Q\left(\gamma, \eta, \pmb{\beta} | \textbf{y}, \textbf{x}, \textbf{w}\right)=\left[\begin{array}{c}
 \sum_{i=1}^{t}  w_{i}\left(\frac{1}{\gamma} - y_{i}^{\eta} e^{\pmb{\beta}^\top \textbf{x}_{i}} \right) \\
 \sum_{i=1}^{t} w_{i}\left(\frac{1}{\eta}+\ln(y_{i})\left(1- \gamma  y_{i}^{\eta} e^{\pmb{\beta}^\top \textbf{x}_{i}}\right)\right)\\
 \sum_{i=1}^{t} w_{i}x_{i1} \Big(1- \gamma y_{i}^{\eta} e^{\pmb{\beta}^\top \textbf{x}_{i}}\Big)\\
\\
\vdots\\
 \sum_{i=1}^{t} w_{i} x_{id} \Big(1- \gamma y_{i}^{\eta} e^{\pmb{\beta}^\top \textbf{x}_{i}}\Big)
\end{array}\right]_{(d+2) \times 1}=\left[\begin{array}{c}
0 \\
0 \\
0 \\
\vdots\\
0
\end{array}\right]_{(d+2) \times 1}.
\end{align*}

\bmsection{The closed form of matrices $\pmb{\Gamma}$ and $\pmb{\Omega}$.}

\renewcommand{\thetable}{A\arabic{table}}

After some mathematical simplifications, we calculate the following symmetric matrices
\begin{align*}
\pmb{\Gamma}&=\mathbb{E}   \left[\begin{array}{cccc}
\sum_{i=1}^{t} w_{i} \frac{\partial^2 l_{i}}{\partial \gamma^2} &\sum_{i=1}^{t} w_{i} \frac{\partial^2 l_{i}}{\partial \gamma \partial \eta}   & \ldots & \sum_{i=1}^{t} w_{i} \frac{\partial^2 l_{i}}{\partial \gamma \partial \beta_{d}}     \\
&\sum_{i=1}^{t} w_{i} \frac{\partial^2 l_{i}}{\partial \eta^2}  & \ldots & \sum_{i=1}^{t} w_{i} \frac{\partial^2 l_{i}}{\partial \eta \partial \beta_{d} }   \\
 & & \ddots & \vdots \\ 
  \text{Sym.} &   &   &\sum_{i=1}^{t} w_{i} \frac{\partial^2 l_{i}}{\partial \beta_{id}^2}  
\end{array}\right]=-\sum_{i=1}^{t} w_{i} e^{\pmb{\beta}^\top \textbf{x}_{i}} \mathbb{E}  \left[\begin{array}{cccc}
\frac{e^{-\pmb{\beta}^\top \textbf{x}_{i}}}{\gamma^2}&  Y_{i}^{\eta} \ln(Y_{i}) & \cdots &  Y_{i}^{\eta} x_{id}  \\
   & \frac{e^{-\pmb{\beta}^\top \textbf{x}_{i}}}{\eta^2}+ \gamma  Y_{i}^{\eta} \ln^2(Y_{i}) & \cdots & \gamma Y_{i}^{\eta} \ln(Y_{i}) x_{id} \\
  &  & \ddots & \vdots \\ 
  \text{Sym.}  &   &  & \gamma Y_{i}^{\eta} x_{id}^2 
\end{array}\right],
\end{align*}
 and
\begin{align*}
&\pmb{\Omega}=\sum_{i=1}^{t} w_{i}^2 \mathbb{E} \left[\begin{array}{cccc}
\left(\frac{1}{\gamma} -  Y_{i}^{\eta} e^{\pmb{\beta}^\top \textbf{x}_{i}}\right)^2& \left(\frac{1}{\gamma} -  Y_{i}^{\eta} e^{\pmb{\beta}^\top \textbf{x}_{i}}\right)  \left(\frac{1}{\eta}+\ln(Y_{i})\Big(1-  \gamma  Y_{i}^{\eta} e^{\pmb{\beta}^\top \textbf{x}_{i}}\Big)\right) & \cdots &  \left(\frac{1}{\gamma} -  Y_{i}^{\eta} e^{\pmb{\beta}^\top \textbf{x}_{i}}\right)\left(x_{id} \Big(1- \gamma Y_{i}^{\eta} e^{\pmb{\beta}^\top \textbf{x}_{i}}\Big)\right)   \\
  & \left(\frac{1}{\eta}+\ln(Y_{i})\Big(1-  \gamma  Y_{i}^{\eta} e^{\pmb{\beta}^\top \textbf{x}_{i}}\Big)\right)^2 & \cdots &  \left(\frac{1}{\eta}+\ln(Y_{i})\Big(1-  \gamma  Y_{i}^{\eta} e^{\pmb{\beta}^\top \textbf{x}_{i}}\Big)\right) \left(x_{id} \Big(1- \gamma Y_{i}^{\eta} e^{\pmb{\beta}^\top \textbf{x}_{i}}\Big)\right)\\
 & & \ddots & \vdots \\ 
 \text{Sym.} & &   &\left(x_{id} \Big(1- \gamma Y_{i}^{\eta} e^{\pmb{\beta}^\top \textbf{x}_{i}}\Big)\right)^2  
\end{array}
\right].
\end{align*}

To calculate the matrices $\pmb{\Gamma}$ and $\pmb{\Omega}$, note that

\begin{align*}
\mathbb{E}(Y^{\eta})&=\frac{1}{\gamma} e^{\beta x}\nonumber\\
\mathbb{E}(Y^{\eta}\ln(Y))&=\frac{-e^{\beta x}\left(\beta x+\ln(\gamma+c-1)\right)}{\gamma \eta}\nonumber\\
\mathbb{E}(Y^{\eta}\ln^2(Y))&=\frac{-e^{\beta x}\left(\ln^2(\gamma)+2\ln(\gamma)(\beta x+c-1)+c^2+2c(\beta x-1)+\beta^2 x^2-2\beta x+\frac{\pi}{6}\right)}{\gamma \eta^2}\nonumber\\
\mathbb{E}\left[\left(\frac{1}{\eta}+\ln(Y)\Big(1-  \gamma  Y^{\eta} e^{\beta x}\Big)\right)^2\right]&=\frac{6+6\ln^2(\gamma)+12\ln(\gamma)(\beta x+c-1)+6c^2+12c(\beta x-1)+6\beta^2 x^2-12\beta x+\pi^2}{6\eta^2} \nonumber\\
\end{align*}
and 
\begin{align*}
\mathbb{E}\left[\frac{1}{\gamma} -  Y^{\eta} e^{\beta x}\right]^2&=\frac{1}{\gamma^2}\nonumber\\
\mathbb{E}\left[\left(\frac{1}{\gamma} -  Y^{\eta} e^{\beta x}\right)  \left(\frac{1}{\eta}+\ln(Y)\left(1-  \gamma  Y^{\eta} e^{\beta x}\right)\right)\right]&=\frac{-\beta x-c-\ln(\gamma)+1}{\gamma \eta}\nonumber\\
\mathbb{E}\left[\left(\frac{1}{\gamma} -  Y^{\eta} e^{\beta x}\right)\left(x \Big(1- \gamma Y^{\eta} e^{\beta x}\Big)\right)\right]&=\frac{x}{\gamma} \nonumber\\
\mathbb{E}\left[ \left(\frac{1}{\eta}+\ln(Y)\Big(1-  \gamma  Y_{i}^{\eta} e^{\beta x_{i}}\Big)\right) \left(x \Big(1- \gamma Y_{i}^{\eta} e^{\beta x_{i}}\Big)\right)\right]&=\frac{-x\left(\beta x+\ln(\gamma)+c-1\right)}{\eta} \nonumber\\
\mathbb{E}\left[\left(x \Big(1- \gamma Y^{\eta} e^{\beta x}\Big)\right)^2 \right]&=x^2  \nonumber\\
\end{align*}
where $c$ is the Euler's constant.

\end{document}